 \documentclass[final,1p,times]{elsarticle}

\usepackage{graphicx}
\usepackage{amssymb}

\usepackage{lineno}

\usepackage{xcolor}

\journal{J. Informetrics}

\begin{document}

\begin{frontmatter}


 

\title{Impact Factors and the Central Limit Theorem: Why Citation Averages Are Scale Dependent}




\ead{ma2529@columbia.edu}

 \author[label1,label2]{Manolis Antonoyiannakis}
 \address[label1]{Department of Applied Physics \& Applied Mathematics, Columbia University, 500 W. 120th St., Mudd 200, New York, NY 10027}
 \address[label2]{American Physical Society, Editorial Office, 1 Research Road, Ridge, NY 11961-2701}

\begin{abstract}

Citation averages, and Impact Factors (IFs) in particular, are sensitive to sample size.
Here, we apply the {\it Central Limit Theorem} to IFs to understand their scale-dependent behavior. 
For a journal of $n$ randomly selected papers from a population of all papers, we expect from the Theorem that its IF  fluctuates around the population average $\mu$, and spans a range of values proportional to $\sigma/\sqrt[]{n}$, where $\sigma^2$ is the variance of the population's citation distribution. 
The $1/\sqrt[]{n}$ dependence has profound implications for IF rankings: The larger a journal, the narrower the range around $\mu$ where its IF lies. 
IF rankings therefore allocate an unfair advantage to smaller journals in the high IF ranks, and to larger journals in the low IF ranks.
As a result, we expect a scale-dependent stratification of journals in IF rankings, whereby small journals occupy the top, middle, {\it and} bottom ranks; mid-sized journals occupy the middle ranks; and very large journals have IFs that  asymptotically approach $\mu$. 
We obtain qualitative and quantitative confirmation of these predictions by analyzing (i) the complete set of 166,498 IF \& journal-size data pairs in the 1997--2016 Journal Citation Reports of Clarivate Analytics, (ii) the top-cited portion of 276,000 physics papers published in 2014--2015, and (iii) the citation distributions of an arbitrarily sampled list of physics journals. 
We conclude that the Central Limit Theorem is
a good predictor of the IF range of actual journals, while sustained deviations from its predictions are a mark of true, non-random, citation impact.
IF rankings are thus misleading unless one compares like-sized journals or adjusts for these effects. 
We propose the $\Phi$ index, a rescaled IF that accounts for size effects, and which can be readily generalized to account also for different citation practices across research fields.  Our methodology applies to other citation averages that are used to compare research fields, university departments or  countries in various types of rankings.

\end{abstract}


\end{frontmatter}



\section{Introduction}
\label{S:1}

What do crime rates, cancer rates, high-school mean test scores, and Impact Factors have in common? They are all manifestations of the Central Limit Theorem, which explains why small populations (cities, schools, or research journals) score more often than one would expect at the top {\it and} bottom places of rankings, while large populations end up in less remarkable positions. But if size affects one's position in a ranking, then rankings of population averages must be misleading. The Impact Factor is an average measure of the citation impact of journals. Therefore, it may seem perfectly justifiable to use it when ranking journals of different sizes, in the same vein we use averages to rank, say, the class size of schools, the GPA's of students, the fuel efficiency of engines, the life expectancy in countries, or the GDP per capita for various countries. However, underlying such comparisons is the tacit admission (De Veaux, Velleman, \& Bock, 2014)
that the distributions being compared are (approximately) symmetric and do not contain outliers (i.e., extreme values)---or if they do, that the sample sizes are large enough to absorb extreme values. If the distributions are highly skewed, with outliers, and especially if the populations are small, then rankings by averages can be misleading, because averages are no longer representative of the distributions. Impact Factors qualify for these caveats. So far, several studies drew attention to the skewness of the citation distribution, or various other features of the Impact Factor, such as the `free' citations to front-matter items of journals, the need to normalize for different citation practices among fields, the citation time windows, the lack of verifiability in the citation counts entering the Impact Factor calculations, the mixing of document types with disparate citabilities (articles versus reviews), etc. 
(Seglen, 1992; 
Seglen, 1997; 
Redner, 1998;
Rossner, Van Epps, \& Hill, 2007;
Adler, Ewing, \& Taylor, 2008;
Radicchi, Fortunato, \& Castellano, 2008;
Wall, 2009;
Fersht, 2009;
Gl{\"a}nzel \& Moed, 2013;
Antonoyiannakis, 2015a;
Antonoyiannakis, 2015b;
San Francisco Declaration on Research Assessment, 2012;
Bornmann \& Leydesdorff, 2017). 
However, little attention has been paid 
(Amin \& Mabe, 2004;
Antonoyiannakis \& Mitra, 2009)
to the effect of journal scale on Impact Factors, which, as we will show, is substantial.

The Impact Factor is defined as 
\begin{equation}
{\rm IF}=\frac{C}{N_{2Y}}=\frac{\sum_{i=1}^{i=N_{2Y}}c_i}{N_{2Y}}, \label{eq:JIF-def}
\end{equation}
where $C$ are the citations received in year $y$ to journal content published in years $y-1,y-2$, and $N_{2Y}$ is the biennial publication count, i.e., the number of citable items (articles and reviews) published in years $y-1,y-2$. As can be verified from the Journal Citation Reports (JCR) of Clarivate Analytics, the annual publication count of journals ranges from a few papers to a few tens of thousands of papers. At the same time, individual papers can collect from zero to a few thousand citations in the JCR year. With a span of 4 orders of magnitude in the numerator and 5 orders of magnitude in the denominator, the IF is a quantity with considerable room for wiggle. 

In this paper, {\it first}, we apply the Central Limit Theorem (the celebrated theorem of statistics) to understand and predict the behavior of Impact Factors. We find that Impact Factor rankings produce a scale-dependent stratification of journals, as follows. (a) Small journals occupy all ranks (top, middle {\it and} bottom); (b) mid-sized journals occupy the middle ranks; and (c) very large journals (``megajournals") converge to a single Impact Factor value---the population mean---almost irrespective of their size. Impact Factors are thus sensitive to journal size, and Impact Factor rankings do not provide a `level playing field,' because size affects a journal's chances to make it in the top, middle, or bottom ranks. {\it Second}, we apply the Central Limit Theorem to arrive at an {\it Impact Factor uncertainty relation:} an expression that limits the expected range of Impact Factor values for a journal as a function of journal size and the citation variance of the population of all published papers. {\it Third}, we confirm our theoretical results, by analyzing 166,498 IF \& journal-size data pairs, the citation-distribution data from 276,000 physics papers, and an arbitrarily sampled list of physics journals. We observe the predicted scale-dependent stratification of journals. We find that the Impact Factor uncertainty relation is a very good predictor of the range of Impact Factors observed in actual journals. And {\it fourth}, we argue that sustained deviation from the expected IF range is a mark of non-random citation impact. We thus propose to normalize IFs with regard to the theoretically expected maximum at a given size (using appropriate offsets), as a scale-independent index of citation impact. 

Why does all this matter? Because statistically problematic comparisons can lead to misguided decisions, and Impact Factor rankings remain in wide use (and abuse) today 
(Gaind, 2018;
Stephan, Veugelers, \& Wang, 2017).

Our analysis shows that Impact Factor comparisons---even for similar fields and document types---for different-sized journals can be misleading. We argue that it is imperative to a seek metrics that are immune from or correct for this effect.



\section{Theoretical Background}
\label{S:2}

\subsection{The Central Limit Theorem for citation averages (i.e., Impact Factors)}
\label{SS:2.1}

The Central Limit Theorem is the fundamental theorem of statistics. In a nutshell, it says that for independent and identically distributed data whose variance is finite, the sampling distribution of any mean becomes more nearly normal (i.e., Gaussian) as the sample size grows (De Veaux, Velleman, \& Bock, 2014). 
The sample mean $\bar{x}_n$ will then approach the population mean $\mu$, {\it in distribution}. More formally, 

\begin{equation}
\lim_{n\to\infty}\Big(\sqrt[]{n} \big(\frac{\bar{x}_n-\mu}{\sigma}\big)\Big) \,{\buildrel d \over =}\, N(0,1) \label{eq:CLT-formal}
\end{equation}
whence  
\begin{equation}
\sigma_n = \frac{\sigma}{\sqrt[]{n}}, \label{eq:CLT}
\end{equation}
where $N(0,1)$ is the normal distribution and the symbol ``d'' in the equality means {\it in distribution}. $\sigma_n$ is the standard deviation of a sampling distribution, $\sigma$ is the standard deviation of the entire population we wish to study (and which is often not known), and $n$  the sample size. So, sample means vary less than individual measurements. (The square of the standard deviation is the {\it variance}.)

The sampling distribution is a notional (imaginary) distribution from a very large number of samples, each one of size $n$, which approaches a normal distribution in the limit of large $n$. In practice, the Central Limit Theorem holds for $n$ as  low as 30, unless there are exceptional circumstances---e.g., when the population distribution is highly skewed---in which case higher values are needed. So, $\sigma_n$ measures how widely the sample means of size $n$ vary around the the population mean $\mu$ (which is approached in the limit of large $n$).

Before we apply the Central Limit Theorem to IFs, let us comment on the assumptions involved. First, our population consists of all papers (articles and reviews) published in all research fields over a number of previous years (normally two for the IF), and the quantity we are going to average over is the number of citations of each paper received in the current year. The population mean $\mu$ is the citation average of all papers, i.e., the IF of the `megajournal' consisting of all papers in all fields. The sampling involves randomly drawing $n$ papers from the population---i.e., forming a `journal'---and calculating their citation average, which is essentially the IF of the random sample (journal). The Central Limit Theorem then tells us that the IF of all random samples (journals) approaches the population mean $\mu$ {\it in distribution} as $n$ becomes large, and describes how the variance of all IFs of $n$-sized journals depends on $n$ and on the population properties ($\sigma, \mu$). Of course, in real journals the papers are not randomly drawn but {\it selected} by editors, board members, and referees who are consciously trying to `bias' the decision process in favor of the better papers for the benefit of their readers. We will revisit this assumption of random samples in \S \ref{S:6} and \S \ref{S:7} once we present our analysis and data.

Let us now examine the two quantities ($\sigma$ and $n$) on which the sample standard deviation ($\sigma_n$) depends in Eq. (\ref{eq:CLT}).

\subsubsection{Variance effects (dependence on $\sigma$)}
\label{SSS:2.1.1}

Equation (\ref{eq:CLT}) shows that the sample variance is proportional to the population variance. High variance (i.e., variability, disparity of values) in the population causes high variance in the sample. This makes sense. For example, imagine that the world's richest and tallest persons simultaneously move into a neighborhood of a population of 1000 people. Because income disparity (variance) among the population is far greater than height disparity, we would expect the income means (averages) of various random samples drawn from the population to vary more than height means. Note that citation `wealth' is very unevenly distributed, like monetary wealth. 

For populations of scientific research papers, the citation distributions have high variances, because the individual papers can be cited from 0 to a few thousand times. Therefore citation means (Impact Factors) have a much higher variance at a given sample size, compared to, say, the height means for adults.

It is the high variance ($\sigma^2$) of citations in populations of scientific papers that makes the Central Limit Theorem relevant in Impact Factor rankings. Had $\sigma$ been 100 times smaller for citation distributions, none of the effects described in this paper would be seen---they would be there, of course, but they would be too small to be of interest and would not interfere with rankings of average metrics. Thus, the multiplier $\sigma$ in the numerator of Eq. (\ref{eq:CLT}) acts as a `switch' that turns on the size effects of the denominator.

\subsubsection{Size effects (dependence on $n$)}
\label{SSS:2.1.2}

The inverse square root dependence of Eq. (\ref{eq:CLT}) with sample size $n$ means that for small journals (small $n$), IFs can fluctuate widely around the population mean $\mu$. Thus, for small journals we expect to see a wide range of IFs, from very low to very high values.  Small journals  will thus dominate the high ranks of Impact Factor values, but also the low ranks! Actually, small journals will cover the entire range of Impact Factor values. With increasing $n$ to medium-sized journals, the fluctuation $\sigma_n$ around the population mean $\mu$ decreases. Therefore, mid-sized journals will not be able to achieve as high Impact Factors as small journals but  they will be spared from really low values too. So, mid-sized journals will do better than small journals in the low ranks but worse in the top ranks. Finally, for large journal sizes $n$, the fluctuation of sample means around the population mean $\mu$ is small, so all Impact Factors of large journals will asymptotically approach $\mu$. Therefore, very large journals have no chance at all to populate any remarkable (i.e., prestigious) ranks. However,  they will  be ranked higher than many small (and a few mid-sized) journals. 

We can codify the above discussion in a simple conceptual diagram. For simplicity, let us use three size classifications, as follows. We classify journals with biennial publication count $n\leq 2000$ as `small;' journals with $2000 < n \leq 10000$ as `mid-sized;' and journals with $n > 10000$ as `large.' We would then expect the scale-dependent stratification of journals in IF rankings that is shown in Table 1. 

By the way, such stratification effects have been reported for other average metrics---e.g., crime statistics, school performances, cancer rates, etc.---and explained in terms of the Central Limit Theorem 
(Wainer \& Zwerling, 2006;
Wainer, 2007;
Gelman \& Nolan, 2002). 
In one spectacular example that made headlines, the Bill and Melinda Gates Foundation ``began funding an effort to encourage the breakup of large schools into smaller schools" since it ``had been noticed that smaller schools were more common among the best-performing schools than one would expect" (De Veaux, Velleman, \& Bock, 2014).
However, small schools were also more likely to be among the worst-performing schools, in agreement with the Central Limit Theorem. Ultimately, the Foundation shifted its emphasis away from the small-schools effort. 

\begin{table}
\centering

\begin{tabular}{rl}
\hline
Impact Factor & Journal Size    \\
\hline
High & Small \\
Moderate & Small | Medium  \\
Average & Small | Medium | Large  \\
Below average & Small | Medium  \\
Low & Small  \\
\hline

\end{tabular}
\caption{Because of the Central Limit Theorem, we expect a scale-dependent stratification of journals in IF rankings.}

\end{table}

\subsubsection{Why is the Central Limit Theorem relevant for Impact Factors? }
\label{SSS:2.1.3}

Journal sizes range typically from 50--50,000 (biennial count, $n$), so the quantity $1/\sqrt[]{n}$ ranges from $10^{-3}$--$10^{-1}$. If the population standard deviation $\sigma$ were no greater than 1, then random fluctuations in Impact Factors (which are a few {\it times} the $\sigma_n$) would be  smaller than 0.5, say, and thus irrelevant for journal rankings (except for very low Impact Factors). But if, as we will show later, $\sigma$ is at least 10, then $\sigma_n$ lies in the range 0.05--1.5, and journal rankings are affected significantly, because random fluctuations lie in the range $\simeq $0.2--5 and are no longer negligible compared to Impact Factors themselves. This is why the Central Limit Theorem is relevant here. (This is a {\it first} justification of the applicability of the Theorem for Impact Factors, see \S \ref{SS:2.1}. More reasons will be provided later.)

\subsection{An uncertainty relation for Impact Factors}
\label{SS:2.2}

Consider the population of  citations in a certain year to all papers published in the previous two years. Imagine that we draw random samples (``journals") of size $n$ from this population, and calculate their citation average, $f(n)$, which for practical purposes is equal to the Impact Factor of the $n$ papers (we ignore from the `free' citations in the numerator, which generally do not play a major role in IF values). Because the sampling distribution of the sample means is normal (i.e., Gaussian), we can expect roughly 68\% of  $f(n)$ values to lie within $\pm \sigma_n$ of  $\mu$, 95\% of  $f(n)$ values to lie within $\pm 2\sigma_n$ of  $\mu$, 99.7\% of  $f(n)$ values to lie within $\pm 3\sigma_n$ of  $\mu$, and 99.99\% of  $f(n)$ values to lie within $\pm 4\sigma_n$ of  $\mu$. So, we can write that for an integer $k$ (where, in practice, $k=3$ or 4), the $f(n)$ values are bounded as
\begin{equation}
\mu - k\sigma_n \leq  f(n) \leq \mu + k\sigma_n. \label{eq:IFUP.1}
\end{equation}
We invoke the Central Limit Theorem, Eq. (\ref{eq:CLT}), to rewrite the above inequality as 
\begin{equation}
\mu - \frac{k\sigma}{\sqrt[]{n}} \leq  f(n) \leq \mu + \frac{k\sigma}{\sqrt[]{n}}. \;\;\;\;\;\;\;\;    {\rm \it  (Impact \, Factor \, Uncertainty \, Relation)} \label{eq:IFUP.2}
\end{equation}
The inequality (\ref{eq:IFUP.2}) says that the IF values that are statistically available to a randomly formed journal range from the theoretical minimum, $f_{min}^{th}(n)$, to the theoretical maximum, $f_{min}^{th}(n)$, which are defined as
\begin{equation}
f_{min}^{th}(n)=\mu - \frac{k\sigma}{\sqrt[]{n}} \;\;\;\;\;\;\; {\rm and}  \;\;\;\;\;\;\; f_{max}^{th}(n)=\mu + \frac{k\sigma}{\sqrt[]{n}}.   \label{eq:fmax-fmin}
\end{equation}
We can recast expression (\ref{eq:IFUP.2}) as 
\begin{equation}
 \Delta f(n) \cdot \sqrt[]{n} \leq k \sigma,  \; {\rm where} \;  \Delta f(n) \equiv | f(n) - \mu |, \label{eq:IFUP.3}
\end{equation}
whence the term {\it uncertainty relation} becomes evident. Indeed, the `uncertainty' $\Delta f(n)$---the range of values of $f(n)$ as measured from $\mu$---multiplied by the square root of the journal (biennial) size $n$ cannot exceed $k\sigma$, statistically speaking. Therefore, for small journals $\Delta f(n)$ is large, while for large journals $\Delta f(n)$ is  small. Again, the expressions (\ref{eq:IFUP.2}) or (\ref{eq:IFUP.3}) hold in a statistical sense---roughly in 99.99\% cases for $k=4$, or at a distance of 4$\sigma_n$ from $\mu$. 

The Impact Factor uncertainty relation has important practical implications, as we discuss below. 

{\it Implication \#1}. Expression (\ref{eq:IFUP.2}) says that Impact Factor uncertainties $\Delta f(n)$ have a maximum value $k \sigma /\sqrt[]{n}$ that is inversely proportional to the square root of  journal size. This scale dependence is rather punitive  for large journals: A 100-fold increase in journal size yields a 10-fold decrease in the range of Impact Factor values, as measured from $\mu$. Therefore, large journals cannot have very high (or very low) Impact Factors. Conversely, small journals have a high range of Impact Factor values. Therefore, small journals  {\it can} reach very high (and very low) Impact Factors. For small $n$ compared to $k^2 \sigma^2$, the term $k\sigma /\sqrt[]{n}$ is large and $\mu$ can be dropped from expression (\ref{eq:IFUP.2}), so the highest Impact Factor is inversely proportional to $\sqrt[]{n}$. 

{\it Implication \#2}. The theoretical maximum $f_{max}^{th}(n)$ increases with the population standard deviation $\sigma$, which is a measure of the disparity (variability) of citations among all papers in the population. 
The theoretical minimum $f_{min}^{th}(n)$ decreases with $\sigma$ by the same amount. The maximum uncertainty $\Delta f_{max}(n) = |f_{max}^{th}(n)-\mu|=k \sigma/\sqrt[]{n}$, and hence the range of IF values, both increase then. That is, as $\sigma$ increases in a population, the range of IF values broadens. 

But what is the population? We have assumed so far that it consists of all papers in all research fields, and this statement is true in a general sense. However, for research fields that cite each other little or not at all, one can claim that they are distinct populations, each with its own $\sigma$ and $\mu$. In this case, expression (\ref{eq:IFUP.2}) says that journals from the population with larger $\sigma$ can reach higher Impact Factors. This is why, for example, mathematics journals have lower IFs compared to chemistry journals (and why normalizing citation averages to account for the different citedness of research fields makes sense and is standard practice in bibliometrics). 

As the readers may have inferred, the parameters $\sigma$ and $\mu$ incorporate everything that makes one population (research field) distinct from another, citations-wise. Differences among fields such as (i) size of reference lists, (ii) chronological distribution of citations, (iii) size of research field, (iv) document types being cited, etc., are all factored in $\sigma$ and $\mu$.

For the remainder of this manuscript, we do not account for different citation practices across research fields---a standard practice in bibliometrics (Moed, 2005, Bornmann \&  Leydesdorff, 2018, Waltman {\it et al}., 2012). This is of course an oversimplification, but to address it here would take us beyond the scope of this paper. We will address this issue in future work.

{\it Implication \#3}. For large enough $n$ there is an Impact Factor minimum, equal to $\mu - k\sigma / \sqrt[]{n}$ (where $f(n)$ has to be nonnegative, of course). That is, $f(n)$ is bounded {\it from below}. 

Let us take a more detailed look at the Impact Factor uncertainty relation--- Eq. (\ref{eq:IFUP.2})---for two limits of interest.

{\it Case I. Small journals}. If there are values $\sigma, \mu$ such that $ n  \ll  k^2 \sigma^2 /  \mu^2  $, then $\mu$ can be left out and Eq. (\ref{eq:IFUP.2}) simplifies to
\begin{equation}
0 \leq f(n) \leq  \frac{k\sigma}{\sqrt[]{n}}, \;\;\; {\rm for} \;\;\; \mu \ll k\sigma /  \sqrt[]{n}.  \label{eq:fmax|small}
\end{equation}So, for small journals the Impact Factor can range from 0 to a maximum value that is inversely proportional to $\sqrt[]{n}$, which can become quite large for small enough size. In other words, small journals are highly volatile, and they will populate all positions in Impact Factor ranks, from the lowest to the highest.

{\it Case II. Very large journals, i.e.,  $ n  \gg  k^2 \sigma^2 /  \mu^2  $.}

Here, expression (\ref{eq:IFUP.2}) reduces to 

\begin{equation}
\mu - \delta \leq  f(n) \leq  \mu +\delta, \;\;\; {\rm where} \; \delta = \frac{k \sigma}{\sqrt[]{n}}\ll   1,   \label{eq:fmax|mega}
\end{equation}
and the Impact Factor asymptotically approaches the population mean, $\mu$. This is both good and bad news for very large journals in Impact Factor rankings: They will neither populate the low ranks nor the high ranks. These journals sample the population, so to speak, so they are stable and insensitive to size effects. Their Impact Factors are bounded from above and below. 

To recap, the Impact Factor uncertainty relation is merely the result of applying the Central Limit Theorem to citation averages, and using standard properties of the normal distribution.

\section{Materials and Methods}
\label{S:2}
\subsection{Approximating $N_{2Y}\approx 2N_Y$ for easier data retrieval}

We collected data on Impact Factors and citable items $N_Y$ from Clarivate Analytics’ Journal Citation Reports (JCR), in the 20-year period 1997--2016. (Both Science Citation Index Expanded (SCIE) and the Social Sciences Citation Index (SSCI) were selected. In the remainder of the paper, unless noted otherwise, all references to JCR include both SCIE and SSCI editions.) The citable items ($N_Y$) data refer to the JCR year: They are the sum of articles and reviews published by a journal in that year. From the original data, we removed those journals whose Impact Factors or citable items were listed as either non-available or zero, as well as duplicate entries. A total of 166,498 journal datapoints were thus obtained. The $N_Y$ values range from 1 to 31,496, while the Impact Factor values range from 0.027 to 187.04. 

For the purposes of this paper we need data on the IF and its denominator, $N_{2Y}$, the biennial publication count in the two years prior to the JCR year. A practical difficulty arises here. While the JCR list IF values and yearly publication counts ($N_Y$) in the JCR year, they do not list $N_{2Y}$ values. To obtain $N_{2Y}$ data we must check each journal individually in the Web of Science---a conceptually trivial but nevertheless cumbersome procedure for tens of thousands of journal datapoints. However, it is reasonable to assume that the publication count does not change appreciably over the 3-year window spanned by $N_Y$ and $N_{2Y}$, and write 
\begin{equation}
N_{2Y} \approx 2N_Y. \label{eq:N2Y=2NY}
\end{equation}

If the approximation (\ref{eq:N2Y=2NY}) holds for all journals, we would be justified to use $2N_Y$ data as a substitute for $N_{2Y}$. It would thus suffice that $N_{2Y}$ be no greater (or smaller) than a few times the product $2N_Y$. That is,  
\begin{equation}
N_{2Y} = z \; 2N_Y, {\rm where} \; 0.1 \lessapprox z \lessapprox 10. \label{eq:N2Y=n 2NY}
\end{equation}

\begin{figure}
\centering
\includegraphics[width=.8\linewidth]{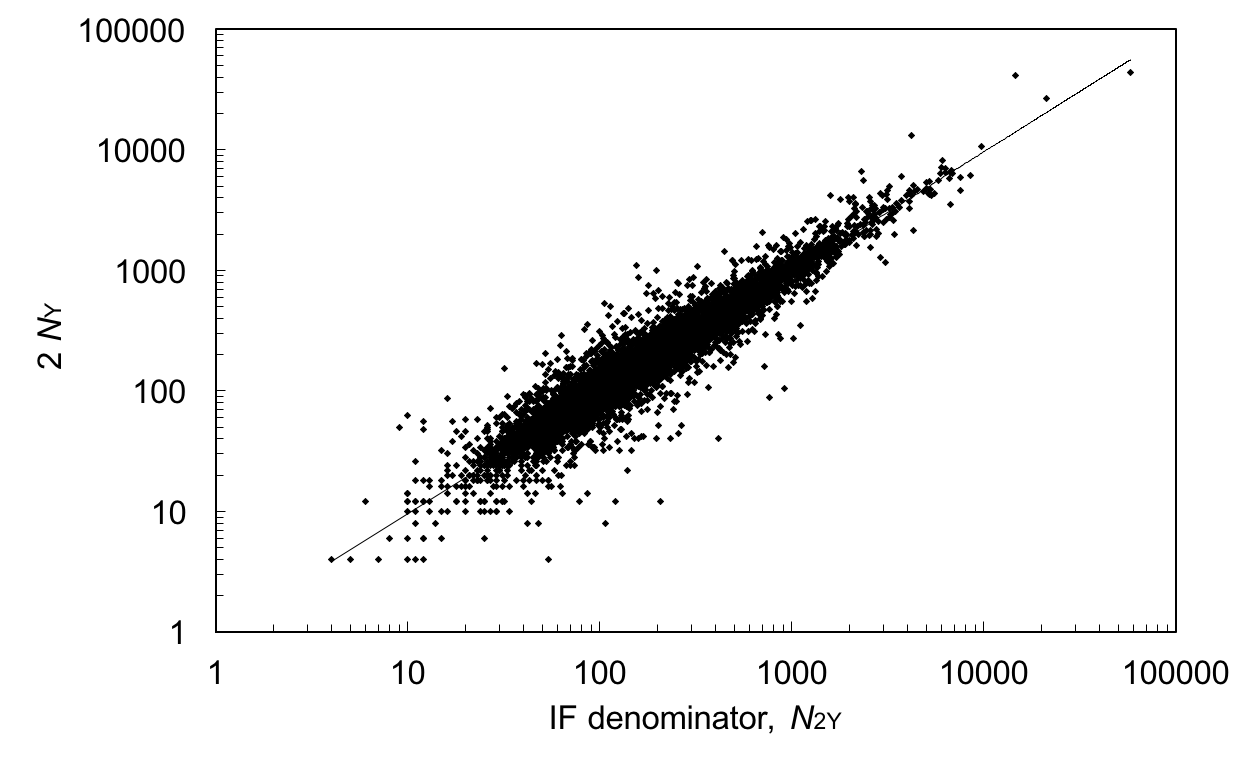}
\caption{How good is the approximation of Eq. (\ref{eq:N2Y=2NY})? We test this from the 2016 JCR SCIE data. The biennial publication count in the two years prior to the JCR year ($N_{2Y}$), is plotted against twice the annual publication count in the JCR year ($2N_{Y}$).}
\label{fig:2NY-N2Y_2016_SCIE}
\end{figure}

We test Eq. (\ref{eq:N2Y=2NY}) for the 2016 JCR year and the 8710 journals in the Science Citation Index Expanded (SCIE) list. As we can see from Fig. \ref{fig:2NY-N2Y_2016_SCIE}, the quantities $2N_Y$ and $N_{2Y}$ are strongly correlated (${\rm slope}=0.96$, $R^2$=0.82, Pearson correlation coefficient = 0.90). Of the 8710 journals, all but 5 (or 99.94\%) satisfy Eq. (\ref{eq:N2Y=n 2NY}), while even for the 5 remaining journals, $z$ remains modest ($z<18$). Therefore, we are justified to use the approximation $N_{2Y} \approx 2N_Y$, provided we are interested in the broad, overall relationship of Impact Factors with journal size. But when we analyze {\it individual} journals, especially with respect to each other (as in {\it ranking}), then we must use $N_{2Y}$.  Certainly, the only reason we may prefer to use $2N_{Y}$ instead of $N_{2Y}$ is the ease of data retrieval from JCR, but where and when necessary, the value $N_{2Y}$ should be used. 

\section{Data Analysis}
\label{S:4}

\subsection{Range of Impact Factor values from citation distribution data}
\label{SS:4.1}

As a first check of the relevance of the Central Limit Theorem---and its underlying sampling distribution---for real distributions, we study the citation distribution of the $\sim$276,000 physics papers (articles and reviews) published in 2014--2015, and cited in 2016, in the Web of Science Core Collection. Since we are interested in the range of IF values, we  calculate how the highest possible citation average depends on sample size for the {\it top-cited} portion of the distribution---that is, for the 2089 papers cited at least 30 times in 2016.

First, we confirm that the citation distribution has finite variance---a key prerequisite for the Central Limit Theorem. Specifically, we find that the probability for a paper among the 2089 papers to be cited more than $n$ times follows a Pareto distribution with a shape parameter $\alpha_P=2.27$, i.e., ${\rm Prob}(x > n) \sim 1/x^{2.27}$.  Because $\alpha_P>2$, the variance is finite. One could easily follow a similar analysis to show that the tails of citation distributions in subjects other than physics also follow a Pareto distribution with finite variance. For example, of the 12633 articles and reviews  published in the subject Information Science and Library Science in 2015--2016, the citations of the 241 papers that were cited at least 10 times in 2017 form a Pareto distribution with a shape parameter $\alpha_P=2.38$. For shape parameter values $\alpha_P \le 2$ the variance of a Pareto distribution is infinite and the Central Limit Theorem does not hold (Newman, 2005).

\begin{figure}
\centering
\includegraphics[width=.8\linewidth]{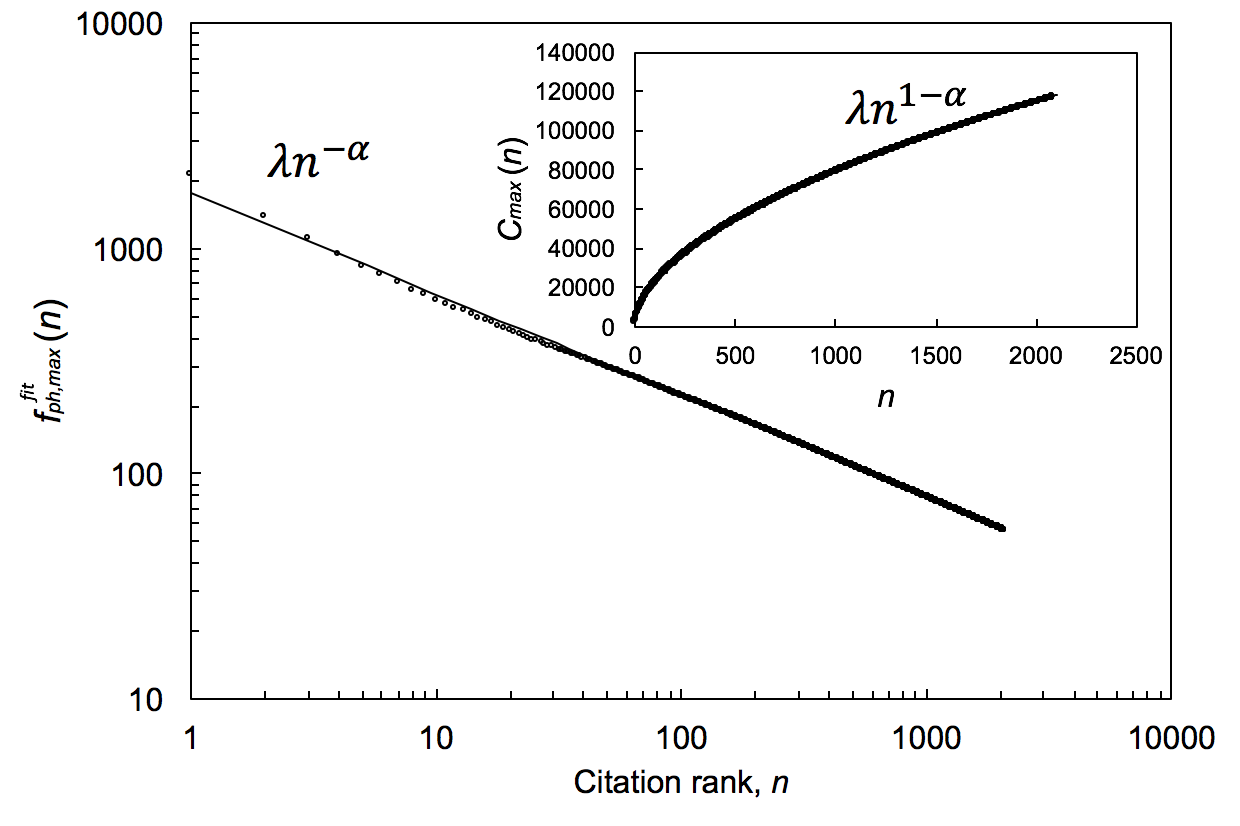}
\caption{Impact Factor, $f_{ph,max}^{fit}(n)$, of the set of $n$ top-cited papers among the 276,000 papers published in physics in 2014--2015, versus citation rank, $n$. Inset: Total citations of $n$ top-cited papers versus $n$. Here, $\lambda=1774$ and $\alpha=0.45$.}
\label{fig:fmax-k_Cmax-k}
\end{figure}

We calculated the total citations, $C_{max}(n)$, as a function of decreasing citation rank, $n$, for the 2089 papers in our set. The dependence of $C_{max}(n)$ on $n$ is found to be (see Fig. \ref{fig:fmax-k_Cmax-k}, inset)
\begin{equation}
C_{max}(n) = \lambda n^{0.55}, \; 1 \leq n \leq 2089, \label{eq:C_max=k^0.56}
\end{equation}
where $\lambda=1774$ is a scale factor in the order of the number of citations received by the most cited paper in this distribution, which in this case is $c_{max}=2121$. Note that $C_{max}(n)$ grows much slower than linearly with $n$. 
The Impact Factor $f_{ph,max}^{fit}(n)$ is defined as the ratio $C_{max}(n)/n$. (We denote results obtained from data fitting with a `fit' superscript, as opposed to results from theory where we use `th'.) So we can write
\begin{equation}
f_{ph,max}^{fit}(n) = \lambda n^{-\alpha}, \label{eq:f_max=k^-0.44}
\end{equation}
with $\alpha=0.45$ and $\lambda=1774$ here. Clearly, $C_{max}(n)$ grows less than linearly with $n$, which results in a size-dependent citation average, $f_{ph,max}^{fit}(n)$. See Fig. \ref{fig:fmax-k_Cmax-k}.

The fact that Eq. (\ref{eq:f_max=k^-0.44}) has been deduced from `only' the top 2089 papers should not distract us from recognizing the generality of the conclusion: Equation (\ref{eq:f_max=k^-0.44}) is in good agreement with the Impact Factor uncertainly relation (\ref{eq:IFUP.2}) and---because the small-journal approximation holds at the high value of  $f_{ph,max}^{fit}(n)$ where Eq. (\ref{eq:f_max=k^-0.44}) is terminated---with its simplified version, Eq. (\ref{eq:fmax|small}). In fact, had we continued the analysis to lower-ranked papers, the exponent in Eq. (\ref{eq:C_max=k^0.56}) would have surely decreased, since the $C_{max}(n)$ curve would cave downward to account for lower-cited papers; consequently, the $\alpha$ value in Eq. (\ref{eq:f_max=k^-0.44}) would approach 0.5. Therefore, the frequency distribution of citation averages for physics articles and reviews agrees with the sampling distribution from the Central Limit Theorem.

To further explore the agreement between actual citation distributions and the sampling distribution, we selected an arbitrarily sampled list of 15 physics journals, and compared  the citation average, $f_{jnl,max}^{fit}(n)$, of their top-$n$ cited papers with the right-hand-side of Eq. (\ref{eq:IFUP.2}), i.e., $f_{max}^{th}(n)=\mu + k\sigma/\sqrt[]{n}$. The journals were Nat. Physics, New J. Phys., Nucl. Phys. B, Phys. Lett. B, Phys. Rev. X, Phys. Rev. Lett.,  Phys. Rev. A, Phys. Rev. B, Phys. Rev. C, Phys. Rev. D, Phys. Rev. E, Phys. Rev. Applied, Phys. Rev. Phys. Ed. Res., Phys. Rev. Acc. Beams, and Rev. Mod. Phys. We obtained the citation distributions for papers published in 2014--2015, cited in 2016. Here, the small-journal approximation does not necessarily hold, so we used the value $\mu=3$ for the population average, a choice that will be justified later. When we plot the quantity $\big(f_{jnl,max}^{fit}(n)-\mu \big)$ against $n$ for the top-25\% cited portion of each journal, we find an $n^{-\alpha}$ behavior. The exponent $\alpha$ ranges from 0.38 to 0.56  for the 15 journals, with a median and mean value both equal to 0.47, in reasonable agreement with the expected value  of 0.50 from the Central Limit Theorem. The median and mean values of $\alpha$ were not particularly sensitive to the specific cutoff point of the top-25\%.

To sum up, the agreement of both the empirical curve Eq. (\ref{eq:f_max=k^-0.44}) from physics papers {\it and} the corresponding curves from our sampled list of 15 physics journals with the Impact Factor uncertainty relation of Eq. (\ref{eq:IFUP.2}) is another confirmation that the Central Limit Theorem applies here. And the key consequence of this applicability is that the range of IF values that are statistically available to a journal of size $n$ scales as $1/\sqrt[]{n}$.

\subsection{Range of Impact Factor values from Impact-Factor \& journal-size data}
\label{SS:4.2}

We have analyzed all 166,498 IF and journal-size data pairs with nonzero values for the Impact Factor and number of annual citable items ($N_Y$) in the Clarivate Analytics’ Journal Citation Reports in the 1997--2016 period. Before we proceed, let us present two important features of Impact Factors and journal sizes, which may not be widely known.

\subsubsection{Small journals are extremely common}
\label{SSS:4.2.1}

\begin{figure}
\centering
\includegraphics[width=.8\linewidth]{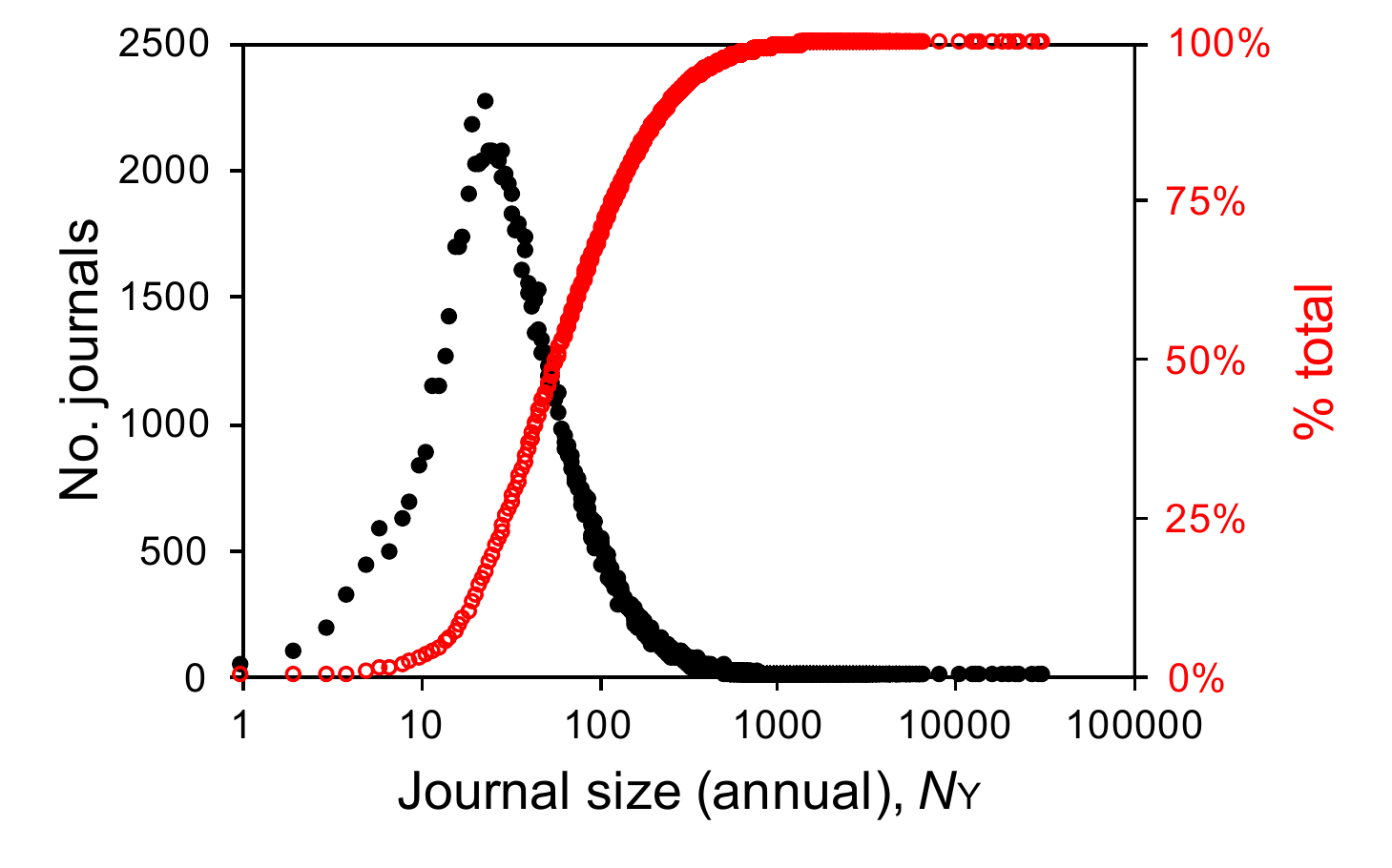}
\caption{Frequency (filled dots) and fraction (hollow dots) of journals with annual publication count $N_Y$. Data for 166,498 journal datapoints in the 1997--2016 JCR.}
\label{fig:freq_jnls-citables}
\end{figure}

In Fig. \ref{fig:freq_jnls-citables} we plot the frequency distribution of journals vs. their annual size, i.e., the number of citable items (articles and reviews) published in the JCR year, $N_Y$. Small journals are extremely common: The most common journal size is 24 citable items per year. 50\% of all journals publish 60 or fewer citable items per year, while 90\% of all journals publish 250 or fewer citable items annually. 

\subsubsection{Most journals have low Impact Factors}
\label{SSS:4.2.2}

\begin{figure}
\centering
\includegraphics[width=.8\linewidth]{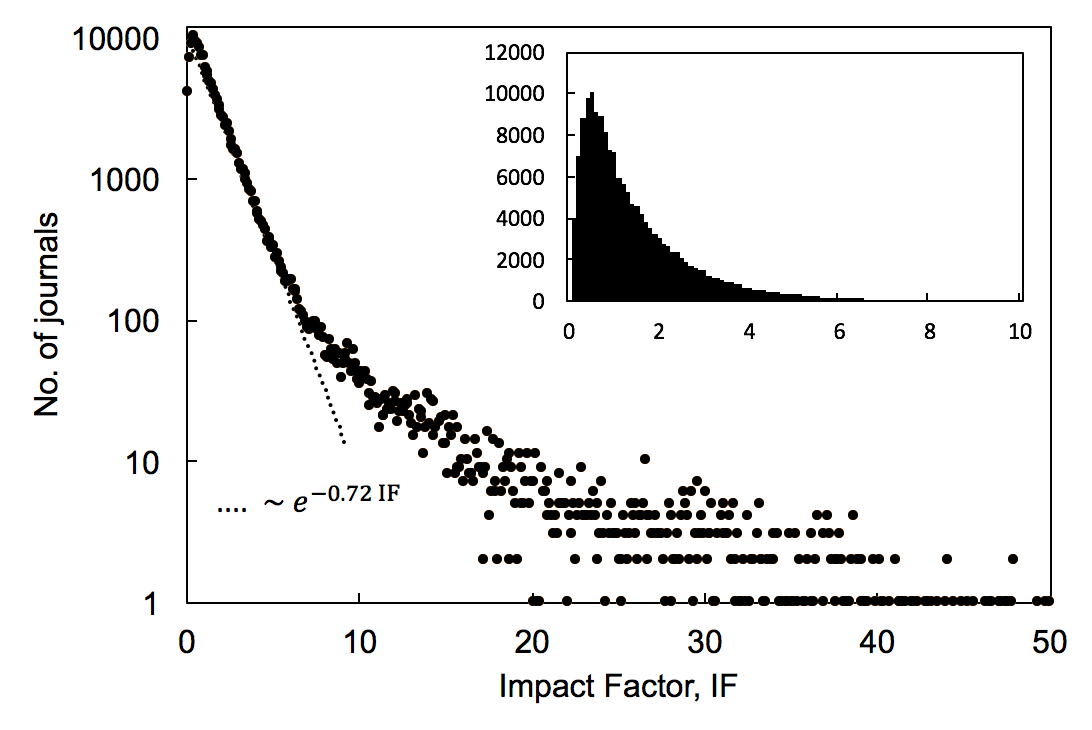}
\caption{Frequency distributions of IF values. The dotted line is an exponential fit, valid ($R^2=0.99$) in the range $0.5 < {\rm IF} < 7$. Inset: Same data but in the range $0 < {\rm IF} < 10$, plotted as a histogram in linear scale. Data for 166,498 journal datapoints in the 1997--2016 JCR. The bin width of ${\rm IF}$ values used in the distribution is 0.1. 
}
\label{fig:freq-JIF}
\end{figure}

In Fig. \ref{fig:freq-JIF} we plot the frequency  distribution of journals vs. their IFs. As is evident from the figure, most IFs are quite low: The most commonly occurring value is 0.5. In the range $0.5 < {\rm IF} < 7$, which covers 85\% of all journals, the frequency distribution can be approximated by an exponentially decreasing function (see dotted line).

We are now ready to analyze Impact-Factor and journal-size data to obtain a boundary curve for Impact Factors. 

\subsubsection{Impact Factor vs. journal size}
\label{SSS:4.2.3}

\begin{figure*}
\centering
\includegraphics[width=11.4cm,height=7.3cm]{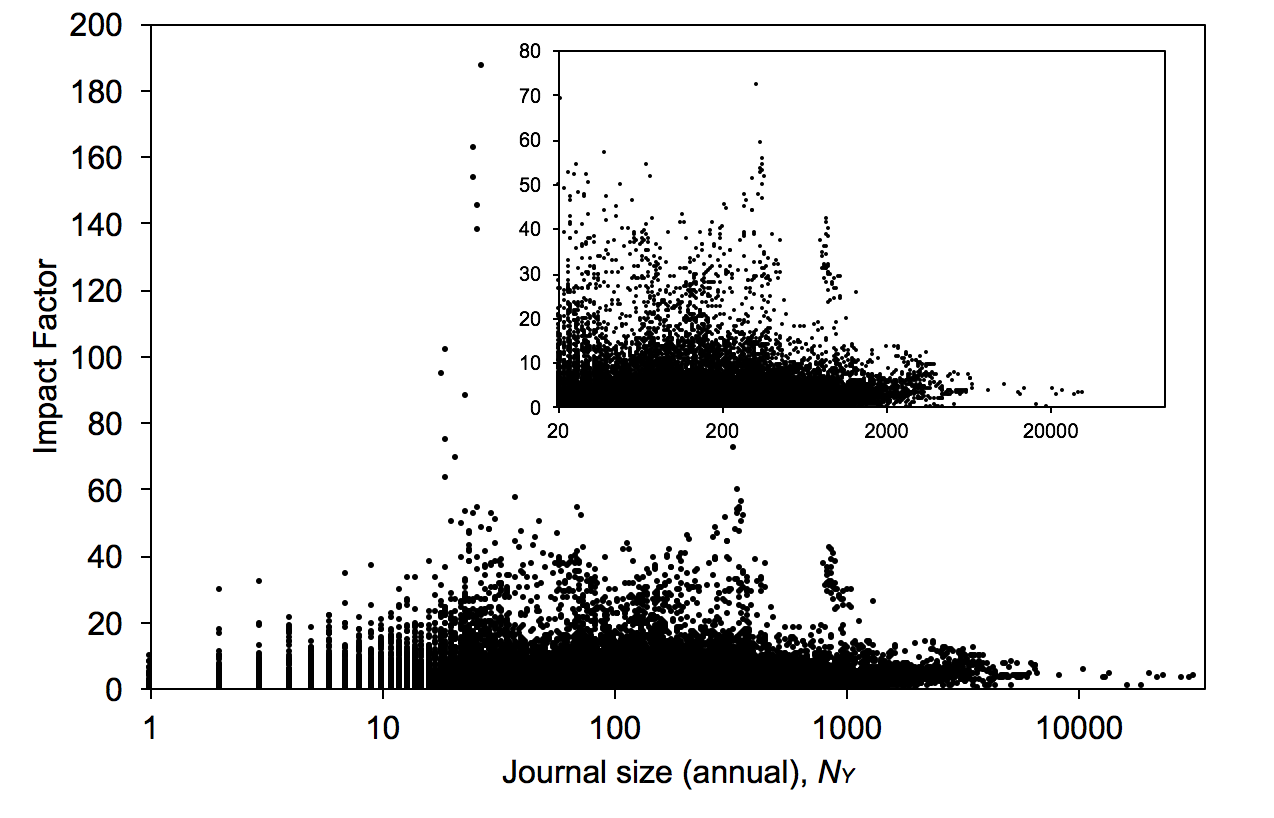}
\caption{IF values versus annual journal size $N_Y$. 166,498 points shown, corresponding to all journals with a nonzero IF and $N_Y$ from 1997--2016. All values shown. Data from Journal Citation Reports, Clarivate Analytics. Inset: Detail of main plot, for $N_Y \geq 30$.}
\label{fig:IF-logNY}
\end{figure*}

In Fig. \ref{fig:IF-logNY}, we plot the Impact Factor versus journal (annual) size for all 166,498 journal datapoints in our set. All values of journal sizes and IFs listed in the 20 years of JCR data are shown in a linear-log plot. (As we noted before, the Central Limit Theorem applies for sample sizes (i.e., biennial journal sizes) that are typically 30 or greater, but for completeness, we show all the data in the figure. More on the range of applicability of the Theorem later.) The first thing to note is that for $n\approx 2 N_Y>60$ (see inset), Fig. \ref{fig:IF-logNY} has the signature appearance of the Central Limit Theorem at work: The data show high variance at small scales, which gradually decreases at higher scales as the datapoints almost converge to a single value at the highest scales. Compare, for instance, with Fig. 3 of Wainer, 2007, 
which shows the age-adjusted kidney-cancer rates in US counties, and which is a well-known manifestation of the Central Limit Theorem.

What happens for $n \le 60$? The abrupt peak of extremely high ($>100$) IF values at biennial size $n \approx 2 N_Y \approx 50$ is a clear break from the gradually diminishing triangular-shaped distribution that is the signature of the Central Limit Theorem. All the datapoints of this peak belong to {\it CA-A Cancer Journal for Clinicians}, a journal whose citation distribution has a very high standard deviation ($\sigma_n \approx  600$, typically). To avoid these ``problematic'' datapoints that challenge the Theorem's validity, we suggest $n=60$ as a conservative lower bound for the biennial sample size of citation distributions. Indeed, for $n>60$, the scatter plot of Fig. \ref{fig:IF-logNY} has largely returned to behavior typical of the Central Limit Theorem. The peaks at $N_Y \approx 350$ and $N_Y \approx 900$ do not deviate drastically from the triangle-shaped distribution. 

A glance at Fig. \ref{fig:IF-logNY} confirms the penalizing effect of journal size on the range of IFs. We observe a global (i.e., large-scale) trend whereby large journals do not have high Impact Factors: Higher IF values tend to occur for smaller ($N_Y<1000$) than larger journals. In broad terms, we observe that of the 166,498 journal datapoints, (a) no journal with $N_Y>2000$ has ${\rm IF}>20$; (b) no journal with $N_Y>1000$ has ${\rm IF}>40$; (c) no journal with $N_Y>500$ has ${\rm IF}>80$; etc. As we zoom in at smaller scales, we notice local irregularities, most notably three local peaks (groups of high-IF datapoints) centered at around $N_Y=25, 350, {\rm and} \; 900$. The first peak ($N_Y=25$) results from very small and selective journals that publish a few mega-cited papers, most notably {\it CA-A Cancer Journal for Clinicians}. The second peak ($N_Y=350$) results from highly selective monodisciplinary journals such as the {\it New England Journal of Medicine}, {\it Lancet}, {\it Chemical Reviews}, {\it Journal of the American Medical Association}, {\it Cell}, {\it Nature Reviews Molecular Cell Biology}, {\it Nature Materials}, {\it Nature Nanotechnology}, etc. And the third peak ($N_Y=900$) is due to highly selective multidisciplinary journals, such as {\it Nature} and {\it Science}.

\begin{figure*}
\centering
\includegraphics[width=11.4cm,height=7.3cm]{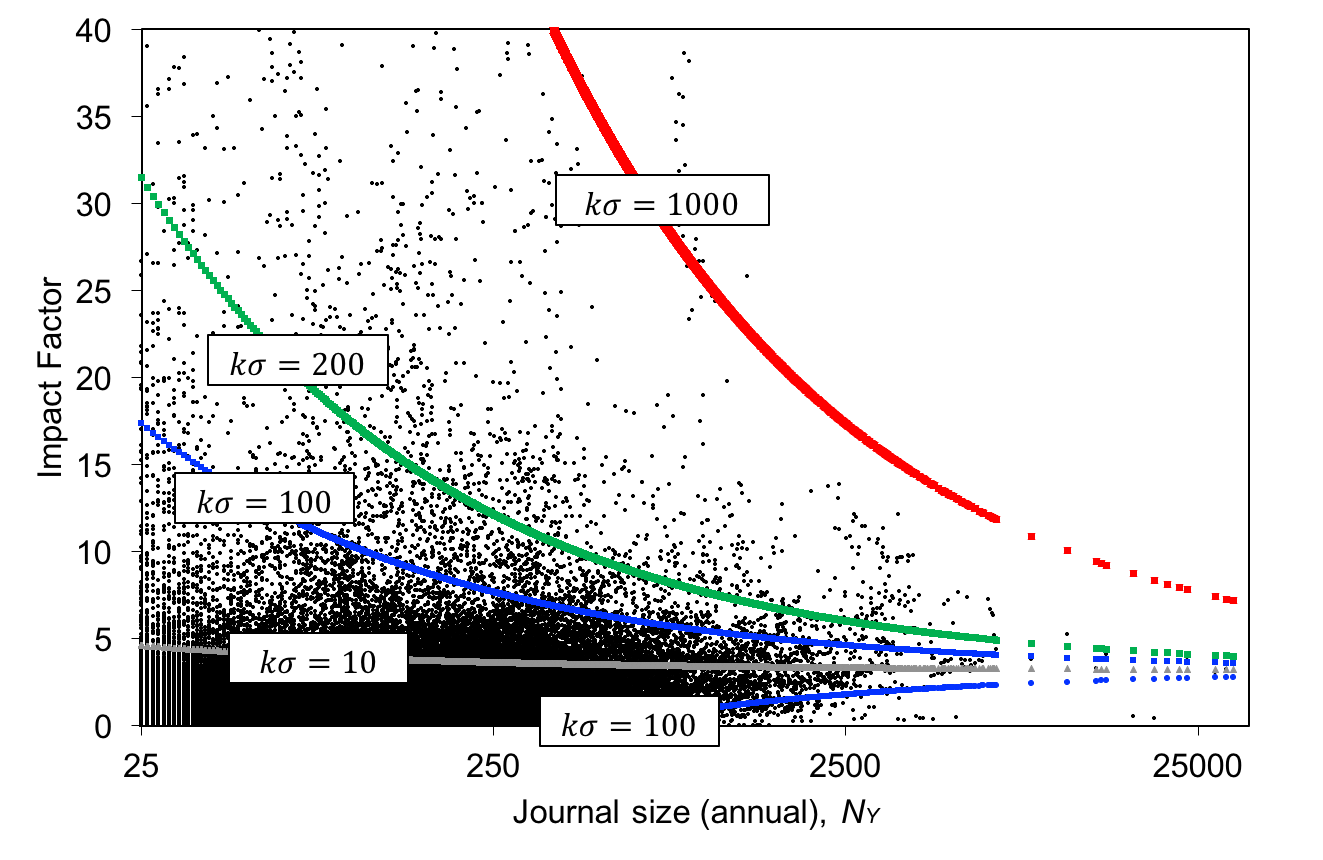}
\caption{Upper and lower bounds of the theoretically expected IFs from the Central Limit Theorem, shown together with actual IF values from JCR data. Four values of $k \sigma$ are shown for upper bounds, one for a lower bound. For high enough values of $k \sigma$, the random fluctuations in the IF described by the Central Limit Theorem are large enough to interfere with IF values.}
\label{fig:IF-ksigma}
\end{figure*}

We are now ready to directly demonstrate the relevance for IF values of the Central Limit Theorem---or the Impact Factor uncertainty relation, Eq. (\ref{eq:IFUP.2}). In Fig. \ref{fig:IF-ksigma} we plot the theoretical upper bound $f_{max}^{th}=\mu + k \sigma/\sqrt[]{n}$,  for $k \sigma = 10, 100, 200, 1000$, and for $\mu=3.2$ (this choice for $\mu$ will be justified later.) We also plot the lower bound $f_{min}^{th}=\mu - k \sigma/\sqrt[]{n}$ for $k \sigma = 100$. In the same plot, we show the IF vs. size data (same data as Fig. \ref{fig:IF-logNY}). Clearly,  for $k \sigma \leq 10$ (i.e., for $\sigma \leq 2.5$ when $k=4$), the Central Limit Theorem is practically irrelevant in IF rankings, because the random variations it considers, which are equal to $k \sigma_n=k \sigma /\sqrt[]{n}$, are much smaller than the vast majority of actual IF values. However, for the other values of  $k \sigma$ listed in the figure, random variations from $\mu$ are clearly comparable to IF values and we can no longer ignore them. And since, as we will shortly explain, the population of scientific papers has $\sigma \geq 15$ and hence $k \sigma \ge 60$, we conclude that accounting for Central Limit Theorem effects in IF rankings is neither a curiosity nor a choice, but a necessity. In fact $k \sigma \approx 100$ is probably more realistic, as we discuss below. 

Why did we choose $\mu=3.2$ and $k \sigma=100$? First, the population mean $\mu$ can be estimated from the Journal Citation Reports (JCR), by summing up all Impact Factor numerators and dividing by all Impact Factor denominators (as these are approximated via Eq. (\ref{eq:N2Y=2NY})). Doing so for the 1997--2016 JCR we obtain $\mu=2.5$. It is reasonable to expect a slight annual inflation of $\mu$, and indeed for the years 2014, 2015, and 2016, we obtain $\mu=3.00, 3.07$, and 3.18, respectively. Now, the presence of $\mu$ is more clearly seen in the high-$n$ limit where the term $k \sigma / \sqrt[]{n}$  is small. And since all but two of the datapoints in this limit  ($n=N_{2Y}=2 N_Y>20000$) correspond to journals that have rapidly grown in size in recent years (namely, RSC Adv., Sci Rep., and {\it PLoS ONE}), we used  the most recent value $\mu=3.18 \approx 3.2$ in Fig. \ref{fig:IF-ksigma}. (We will comment on the two outlier datapoints shortly.) Estimating the population standard deviation $\sigma$ for the 4--5 million articles and reviews published in a two-year period is a little trickier. Our approach is to use the Central Limit Theorem itself to estimate $\sigma$, by drawing a number of randomly selected samples of (approximately) equal size $n$,  calculate the citation mean ${\bar x}_n$ of each sample, calculate $\sigma_n$ from the ${\bar x}_n$ values, and obtain $\sigma$ from Eq. (\ref{eq:CLT}). We performed several tests of this kind from the Web of Science. One practical difficulty was that the ordering of articles in Web of Science searches is not exactly random (even if one orders by date or author name), and in addition, only 100,000 items can be accessed at any time in a search. To avoid spurious effects we performed several types of selections that we tried to make as random as possible, by searching for random ranges of author identifiers (ORCID), or for generic author last names, or for author names starting with a sequence of letters, or for a number of generic words in a topic search, etc. The range of values for $\sigma$ we obtained in this way was 17--148, with a median of 23. For the purposes of this study, we will use the estimate $\sigma  \approx 25$ and we will also take $k=4$---that is, we include random variations within $\pm 4 \sigma_n$ from $\mu$, or at the 99.99\% level---which makes $k \sigma  \approx 100$. The value $\sigma=25$ seems conservative and is in accord with our experience from analyzing journal citation datasets. But even if $\sigma$ were as low as 10, the Central Limit Theorem effects would be relevant in IF rankings, as we discussed above. 

\begin{figure*}
\centering
\includegraphics[width=11.4cm,height=7.3cm]{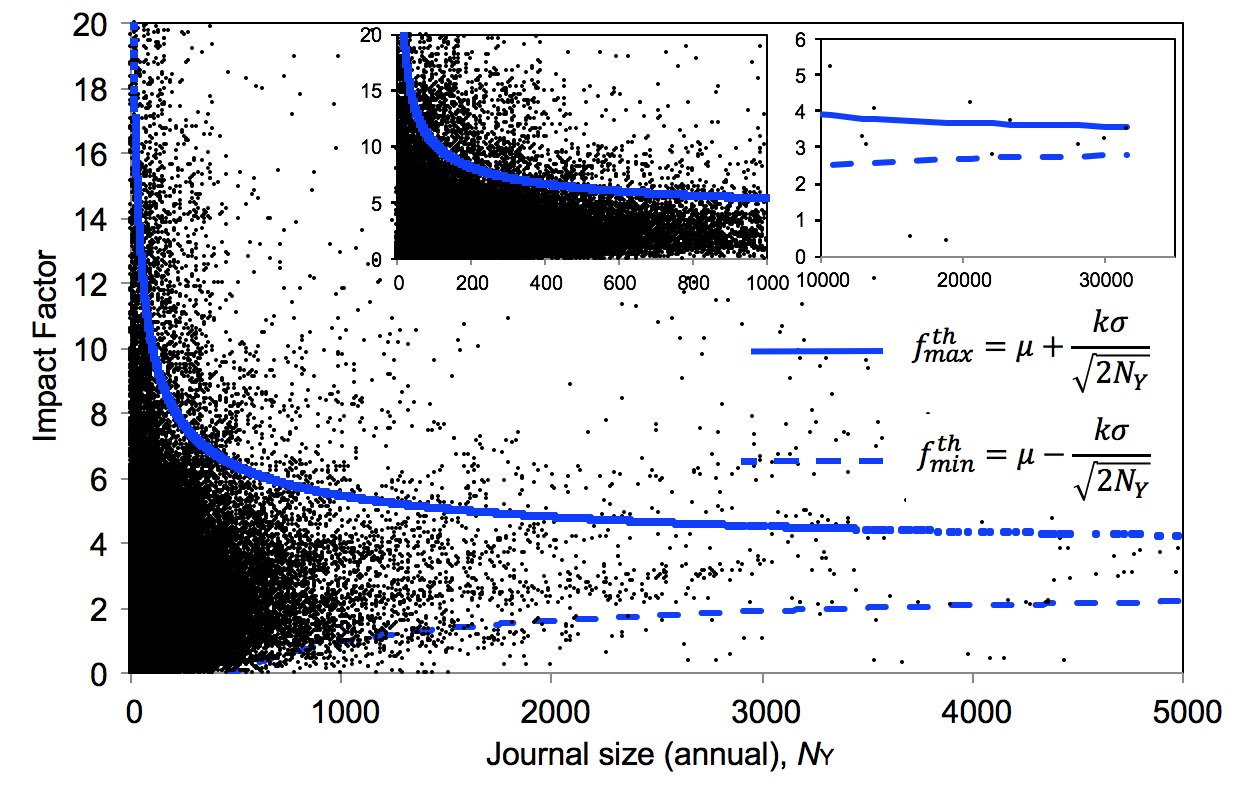}
\caption{IF values versus annual journal size $N_Y$. 166,498 datapoints shown, corresponding to all journals with a nonzero IF and $N_Y$ from 1997--2016. Data from Journal Citation Reports, Clarivate Analytics. Left inset: Detail of main plot, for $N_Y \leq 1000$. Right inset: $N_Y$ values from 10000--35000. The solid and dashed lines are the theoretical maximum and minimum, respectively, from the Central Limit Theorem---see Eq. (\ref{eq:IFUP.2}) The parameters used were $k \sigma= 100$ and $\mu=3.2$.}
\label{fig:JIF-NY_alljnls_clean}
\end{figure*}

 With the parameter choices $k \sigma= 100$ and $\mu=3.2$ at hand, let us now revisit the IF vs. size plot. See Fig. \ref{fig:JIF-NY_alljnls_clean}. The solid and dashed curves are the theoretical maximum and minimum, respectively, from the Impact Factor uncertainty relation, while the insets show the low-$n$ and high-$n$ limit behavior. We observe that the theoretical curves $f_{max}^{th}(n)$ and $f_{min}^{th}(n)$ envelope the IF data and capture the general trend of IF behavior. From the high range of IF values for small journal sizes, to the gradually diminishing IF range with increasing size, to the concentration of IF values within a narrow band in the long-size limit, the theoretical curves capture well what happens with real journals. Actually, 98.1\% of the journals with $N_Y>25$ have IFs within the range [$f_{min}^{th}(n), f_{max}^{th}(n)$]. So there is a good agreement, both qualitatively and quantitatively, between theory (Central Limit Theorem) and `experiment' (IF data). 
 
 Note, incidentally, the two very low (outlier) values of IF, equal to 0.5 and 0.4 for $N_Y \simeq$ 16,000 and $N_Y \simeq$ 19,000 respectively. These datapoints belong to the 2004 and 2005 IF values for {\it Lecture Notes in Computer Science}. Because this is a specialized title that publishes conference proceedings, it corresponds to a distinct population with different citation features (different $\sigma$ and $\mu$) than other large journals--- see the {\it implication \#2} in \S \ref{SS:2.2}. Indeed, this journal is now classified in the Web of Science as Book Series and does not receive an IF value. So it is a special case and does not invalidate the uncertainly relation (\ref{eq:IFUP.2}).

\begin{figure*}
\centering
\includegraphics[width=11.4cm,height=7.3cm]{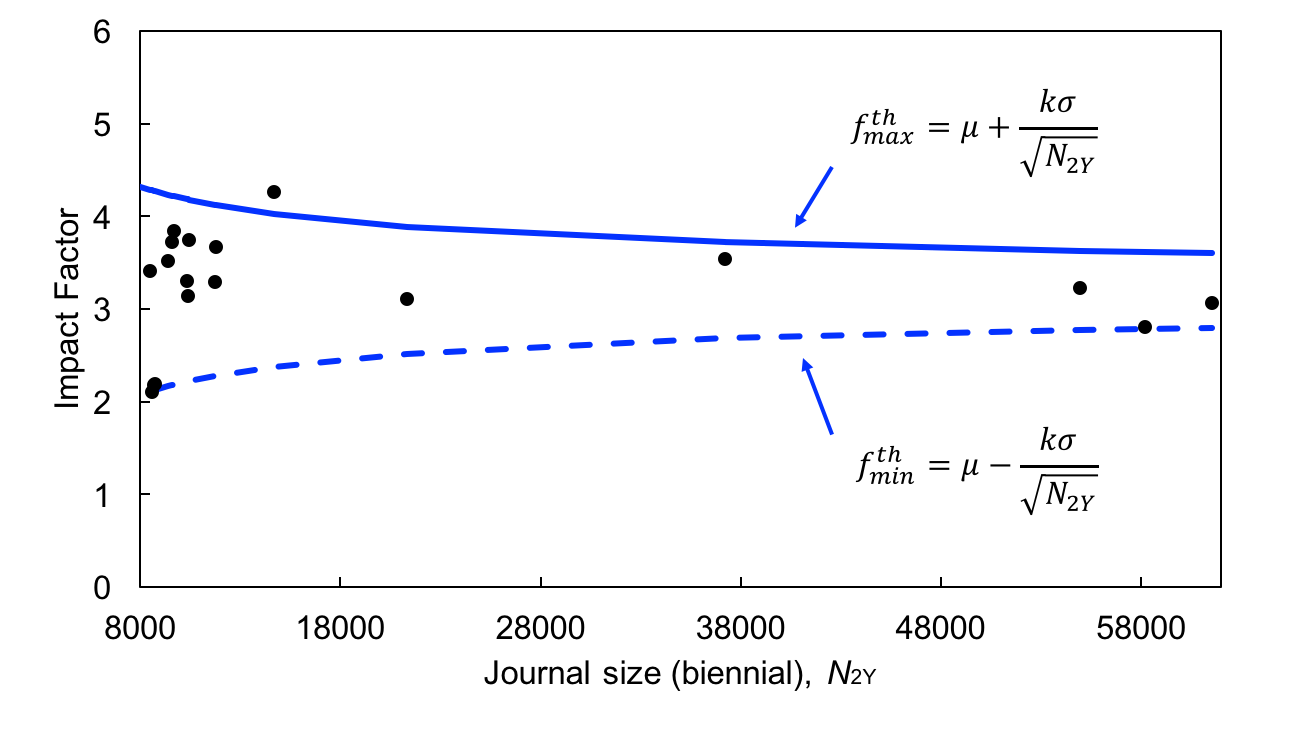}
\caption{IF values versus biennial journal size $N_{2Y}$ for large journals ($N_{2Y}>8000$) from the JCR years 2013--2016. The solid and dashed lines are the theoretical maximum and minimum, respectively, from the Central Limit Theorem---see Eq. (\ref{eq:IFUP.2}) The parameters used were $k \sigma= 100$ and $\mu=3.2$.}
\label{fig:IF-NY-megajournals}
\end{figure*}

While in Fig. \ref{fig:JIF-NY_alljnls_clean} we used the approximation $N_{2Y} \approx 2 N_Y$ for the IF denominator, some of the very large journals had their annual size fluctuate considerably over the last few years (particularly {\it Scientific Reports} and {\it PLoS ONE}). In order to more accurately test the applicability of the Central Limit Theorem in the high-$n$ limit behavior of IF data, we plot in Fig. \ref{fig:IF-NY-megajournals} the IF vs. its exact denominator $N_{2Y}$ for the journals whose biennial size exceeds 8000 in the JCR years 2013--2016. These journals are {\it Applied Physics Letters}, {\it Journal of Applied Physics}, {\it Physical Review B}, {\it RSC Advances}, {\it Scientific Reports}, and {\it PLoS ONE}. Evidently, 17 out of 18 datapoints are within the range [$f_{min}^{th}(n), f_{max}^{th}(n)$], in full confirmation of the Central Limit Theorem. 

To those readers wondering whether ``predatory'' journals appear in Fig. \ref{fig:JIF-NY_alljnls_clean} or Fig. \ref{fig:IF-NY-megajournals}, the answer is negative: Predatory journals, which publish papers for money in an open access model with no regard for the quality or validity of the papers they publish, are generally flagged by gatekeepers at Clarivate Analytics and do not make it in the JCR. Certainly all the megajournals in Fig. \ref{fig:IF-NY-megajournals} are legitimate, respectable journals.

\begin{figure*}
\centering
\includegraphics[width=11.4cm,height=7.3cm]{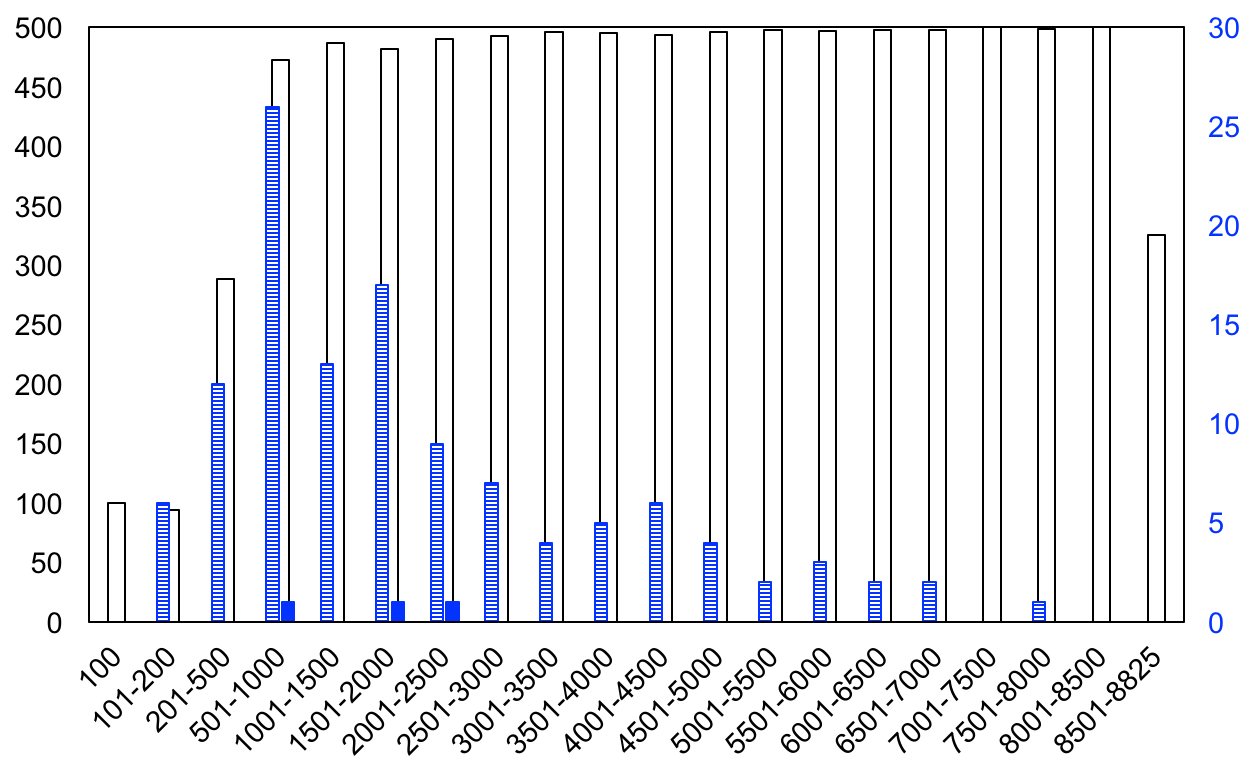}
\caption{Scale-dependent stratification of journals in IF rankings. On the y-axes we plot the frequency of three categories of journals in the ranked positions shown in the x-axis, for the 8825 journals in the 2016 JCR Science Citation Index Expanded list (SCIE). The three journal categories are: small (clear columns; values in primary y-axis on the left), mid-sized (striped blue; values in secondary y-axis on the right), and large (filled blue; values in secondary y-axis), as explained in \S \ref{SSS:2.1.2}. High IF journals appear in top ranks (top 100, 101-200, etc.), low IF journals in the bottom ranks (8001-8500, etc.). This bar plot fully confirms Table 1.}
\label{fig:IFrank-frequency}
\end{figure*}

We predicted earlier (\S \ref{SSS:2.1.2}) a scale-dependent stratification of journals in IF rankings, and this is indeed what we observe.
Again, the (simple) classification system we use is that journals with biennial publication count $n\leq 2000$ are `small'; journals with $2000 < n \leq 10000$ are `mid-sized'; and journals with $n > 10000$ are `large'. 
For example, if we rank by IF the 8825 journals in the 2016 JCR Science Citation Index Expanded list (SCIE), we find that all the top 100 ranks, as well as all the bottom 1315 ranks, are occupied by small journals. The highest rank occupied by a mid-sized journal is \#109, and the lowest rank is \#7510. Consistent with the Central Limit Theorem, the 3 very large journals appear in the rather unremarkable ranks \#973, \#1888, and \#2264. Among the top 500 ranks, 482 slots are taken by small journals and 18 by mid-sized journals. Among the top 1000 ranks, 955 slots are  taken by small journals, 44 slots by mid-sized journals, and 1 slot by a large journal. By rank 5000, 90\% of the mid-sized journals have appeared, as well as all 3 large journals. Finally, among the bottom 3825 ranks (i.e., ranks 5001--8825), 3815 slots are taken by small journals, and 10 by mid-sized journals. The agreement between the observed (Fig. \ref{fig:IFrank-frequency}) and expected (Table 1) stratification of journals in IF rankings is yet another justification of the applicability of the Central Limit Theorem here. Even the asymmetrical position of the large journals and mid-sized journals towards the left of the center of the distribution makes sense: Recall that larger journals approach $\mu$ in the high-$n$ limit while small journals can `fall' all the way to the bottom, and that there are many more low-cited than highly-cited papers for small journals to ``sample."

Some readers may wonder whether the Impact Factor uncertainty relation---and its underlying normal distribution---can account for the IF variations of {\it particular} journals from year to year. That is, given a certain journal, is its annual IF variation smaller than (or comparable to) the random variations expected by the Central Limit Theorem? The answer is positive. In 97.4\% of the 13,099 unique journal titles in our 1997--2016 JCR list whose IF varied anytime but its title remained unchanged in that 20-year period, the maximum uncertainty $|f_{max}^{th}(n)-\mu|= k \sigma/\sqrt[]{n}$ expected from the Central Limit Theorem is greater than the largest IF-variation observed. (However, see also \S \ref{S:6}, where we discuss cases where the annual IF variations are {\it much smaller} than what the Theorem predicts.)

But if IF values fluctuate due to random effects, can we use error bars or confidence intervals on the mean (Impact Factor) to describe these fluctuations? The short answer is {\it yes} for the sampling distribution but {\it no} for the citation distributions of actual journals.  While approximate confidence intervals for means can be constructed even for non-normally distributed data, the use of confidence intervals is based on the premise of a {\it random} variable,  and actual journals are certainly not random samples of papers drawn from the wider population. The sampling distribution, on the other hand, which we have used to estimate IF ranges based on the Central Limit Theorem, pertains to a random variable and we are thus allowed to calculate a standard error or a confidence interval on the mean $f(n)$ of \S \ref{SS:2.2}.  Fundamentally, this is equivalent to our approach in Eq. (\ref{eq:IFUP.2}). We have simply chosen to focus not on $f(n)$ and its standard error, $\sigma/\sqrt[]{n}$, but on the range of $f(n)$ values,  $[f_{min}^{th}(n),f_{max}^{th}(n)]$, which we have found more informative.

\subsubsection{Maximum Impact Factor values from Impact-Factor \& journal-size data}
\label{SSS:4.2.4}

\begin{figure}
\centering
\includegraphics[width=.8\linewidth]{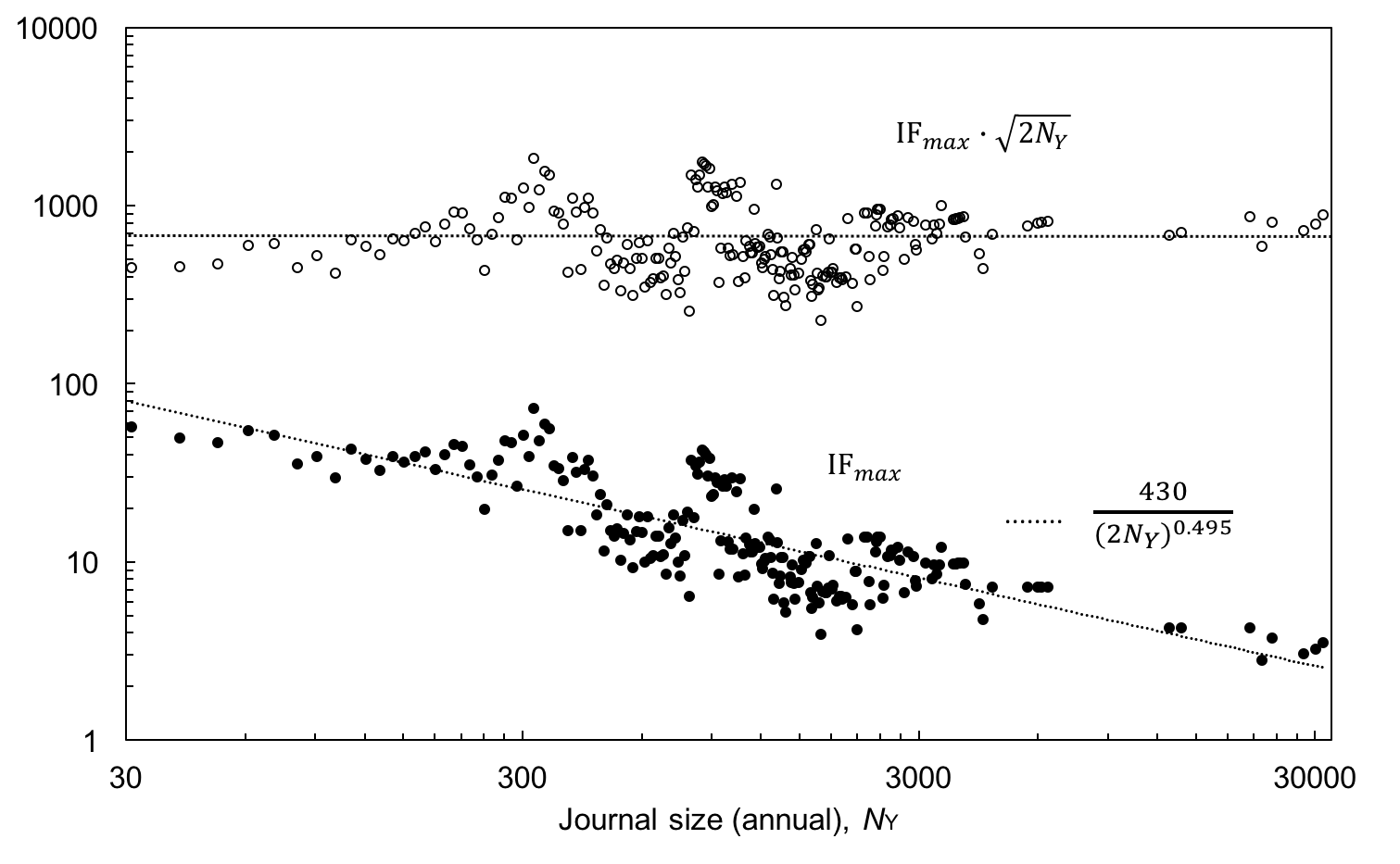}
\caption{Maximum Impact Factor data, ${\rm IF}_{max}$ (filled dots) for each annual journal size, $N_Y$, in the Journal Citation Reports from 1997--2016. The data is drawn from Fig. \ref{fig:JIF-NY_alljnls_clean}, with binning as described in the text. The dotted line (best fit, $R^2=0.64$) is Eq. (\ref{eq:f_fit}) with $\alpha=0.495$ and $\lambda=430$. Also shown (hollow dots) is the product ${\rm IF}_{max}\; \cdot \; \sqrt[]{n}$, which, evidently, does not depend on journal size. ($n=2 N_Y$)}
\label{fig:JIF_max-NY_loglog}
\end{figure}

Let us now check whether the maximum IF values for actual journals have the $1/\sqrt[]{n}$ scale-dependent behavior expected from the Central Limit Theorem. To do that, we extract the dependence of ${\rm IF}_{max}(n)$ on $N_Y$ from data (recall $N_Y=n/2$). So we bin the journal size data of Fig. \ref{fig:JIF-NY_alljnls_clean} in groups (bins) of 10 citable items each ($N_Y=$1--10, 11--20, etc.). For the journals in each bin, we calculate ${\rm IF}_{max}$ and plot it against $N_Y$ in Fig. \ref{fig:JIF_max-NY_loglog} (filled dots). The global downward trend is clearly visible: The journal size, $N_Y$, has an adverse effect on ${\rm IF}_{max}$. Also shown in Fig. \ref{fig:JIF_max-NY_loglog} and is a best-fit curve calculated from the binned data
\begin{equation}
{\rm IF}_{max}^{fit}(n) = \lambda n^{-\alpha}, \;\; n=2N_Y, \label{eq:f_fit}
\end{equation}
where $\alpha=0.495$ and $\lambda=430$.
The details of binning have some effect on the exponent of Eq. (\ref{eq:f_fit}). Clearly, the bin size should not be too large, as it can affect what we are trying to measure, which is size-dependent.   Since the expression (\ref{eq:f_fit}) has an exponent of $0.495\approx 0.5$, we also plot the product (${\rm IF}_{max} \cdot \sqrt[]{n}$) in Fig. \ref{fig:JIF_max-NY_loglog} (hollow dots). Evidently, this product is independent of $N_Y$ (horizontal dotted line), which means that its variance is size-independent (as opposed to the variance of Impact Factors). 

Equation (\ref{eq:f_fit}) confirms the uncertainly relation (\ref{eq:IFUP.2}) from Impact Factor data, just like Eq. (\ref{eq:f_max=k^-0.44}) did from citation data from physics papers. Once again, we observe that the frequency distribution of actual citation averages (IF data) agrees with the (notional) sampling distribution from the Central Limit Theorem, $f_{max}^{th}(n)$. We have thus identified, for the 166,498 journal datapoints in our set, a global boundary curve for Impact Factors as a function of journal size, in the form of Eq. (\ref{eq:fmax|small}), or, more generally, Eq. (\ref{eq:IFUP.2}).

As by-product of the data fitting process in Fig. \ref{fig:JIF_max-NY_loglog}, we observe that the $1/\sqrt[]{n}$ behavior of ${\rm IF}_{max}^{fit}(n)$ appears to hold all the way to large sizes. This indicates that the $k \sigma$ term in the IF uncertainty relation is actually greater than we have assumed, otherwise the constant term $\mu$ would dominate at these sizes. Indeed, the coefficient $\lambda=430$ of Eq. (\ref{eq:f_fit}) provides a measure of $k \sigma$ that is greater than the value 100 that we have so far assumed. This indication for a population variance higher the we assumed ($\sigma \approx 100$ as opposed to $\sigma \approx 25$, for $k=4$) only enhances the effects of the Central Limit Theorem in Impact Factor rankings.

\section{How the Central Limit Theorem tips the balance in IF rankings}
\label{S:5}

The conclusion from the above discussion is unequivocal. The Central Limit Theorem---and its corollary for scholarly journals, the Impact Factor uncertainty relation of Eq. (\ref{eq:IFUP.2})---is an essential tool for explaining and quantifying the scale-dependent behavior of IFs. 

The implications of the $1/\sqrt[]{n}$ dependence of Eq. (\ref{eq:IFUP.2}) for Impact Factors are remarkable. First, the very notion of a constraint implies an unfair advantage: If two athletes are subjected to different constraints for how high they can score, then surely we cannot speak of a level playing field. Likewise, for journals, the range of IF values that are statistically available to different-sized journals depends on their relative size. So the effect of size interferes with the quantity being measured (citation impact) and unless removed, it will  ``tip the balance." Second, the rapidly decaying form of $1/\sqrt[]{n}$ means that the advantage is strongest for small journals, which can thus reach very high Impact Factors. This effect is much stronger than has been previously reported 
(Amin, \& Mabe, 2004). 
It was anticipated though not fully analyzed in our previous work
(Antonoyiannakis \& Mitra, 2009; Antonoyiannakis, 2008). 
Indeed, the ``citation density curve" for Phys. Rev. Letters in 
Antonoyiannakis \& Mitra, 2009,
which is identical to our $ f_{max}^{fit}(n) \sim \lambda n^{-\alpha}$ here, has approximately a $1/\sqrt[]{n}$ dependence on rank $n$, a behavior we had then observed also for other journals. At the time, we had found this a curious feature, but  we now understand its significance. 

But another unfair advantage works also in favor of very large journals when compared to low-IF small journals, since their IFs are bounded {\it from below}---see Eq. (\ref{eq:fmax|mega}). 
First, very large journals (for which $k \sigma/\sqrt[]{n}<1$) will never have very low IFs, i.e., much smaller than $\mu=3.2$. Second, while among the small journals there are many high-IF titles (${\rm IF} \gg \mu $), there are many {\it more} low-IF journals (${\rm IF} < \mu $), and they dominate the statistical correlation of IF with journal size. (This is also why in Fig. \ref{fig:IFrank-frequency} large journals are asymmetrically positioned to the left of the center.) 
Why this asymmetry? For two reasons. First, citation distributions are highly skewed: Low-cited papers are abundant while highly-cited papers are scarce. A random allocation of papers into journals (in the spirit of the Central Limit Theorem) is thus more likely to yield low-IF than high-IF journals. Second (and going beyond the Central Limit Theorem), the stratification of journals by prestige causes the already prestigious journals to disproportionately attract the few highly cited papers, which limits the number of high-IF journals even further.
This behavior is consistent with the findings of 
Rousseau \& Van Hooydonk, 1996,
who report a positive correlation between journal production (i.e., journal size) and Impact Factor for non-review journals grouped in bins of IF values from $0 \leq {\rm IF} \leq 4$. It is also broadly consistent with Huang, 2016,
who reports a positive correlation between article number (size) and IF for scholarly journals in the various subject categories of JCR. (Huang also reports that the correlation is obscure, i.e., not evident, in a direct plot of journal size vs. IF, but because he looks at a very small region of the full spectrum---namely, ${\rm IF} < 5$ and ${\rm size} < 400$---this is not surprising; compare to the full spectrum of IF and journal size values of Fig. \ref{fig:IF-logNY}. Other concerns with the Huang paper were raised by Rousseau, 2016.) 
On a similar note, Fang {\it et al.}, 2018, find that higher IF journals ``publish more papers than expected"---for example, ``journals in the top IF quartile publish 44\% of all SCI papers," in agreement with our analysis: Since most journals are small (see Section 4.2.1) and have low IFs (Section 4.2.2), the bottom quartile of IFs must be populated by small journals, which would therefore account for fewer than 25\% of all papers. Indeed, when we analyze our data in a similar fashion as Fang {\it et al.}, we find that the journals in the top quartile of IF values publish 46\% of the papers. Clearly, the combined effects of the Central Limit Theorem (large journals have ``decent" IFs that approach $\mu$) and the high skewness of citation distributions (plenty of low-cited papers but few highly-cited papers) , exacerbated by the tendency of highly cited papers to congregate in more prestigious journals,
explains the behavior seen by Rousseau \& Van Hooydonk, 1996, Huang, 2016, and Fang {\it et al.}, 2018.

Our findings are also in agreement with 
Katz, 2000, 
who studied the effect of scale of  scientific communities on the citation impact of papers, and reported a power-law behavior that is, in most but not all cases, in favor of larger scientific communities, which he attributed not to the Central Limit Theorem but to the Matthew effect in science (``the rich get richer") 
(Merton, 1968).
In a similar vein, 
van Raan, 2008, 
studied the 100 largest European research universities and reported a size-dependent cumulative advantage of the correlation between the number of citations and number of publications. A closer look at the van Raan study 
reveals that the advantage (power exponent $> 1$) for larger universities decreases as one moves from the bottom- to the top-performing universities, and in fact becomes a {\it dis}advantage (power exponent $< 1$) for the top 10\% performing universities. This behavior is in qualitative agreement with the Impact Factor uncertainty relation of Eq. (\ref{eq:IFUP.2}): The top-performing universities (`samples') have higher citation averages and are thus closer to the $f_{max}^{th}$ curve, which decreases with size (since it has negative slope, see the right-hand side of Eq. (\ref{eq:IFUP.2})); while for the bottom 10\% universities there is gain to be had in increasing in size, because they are closer to the $f_{min}^{th}$ curve that increases with size (has positive slope, see the left-hand side of Eq. (\ref{eq:IFUP.2})). Paraphrasing the van Raan study, we could say that, statistically speaking, low-IF journals will gain by increasing their size, while high-IF journals will lose. As we see, it is not necessary to attribute such scale-dependent effects to the Matthew effect, while the Central Limit Theorem provides a more nuanced, elegant, and quantitatively-minded explanation.


Crespo {\it et al.}, 2012,
proposed a method to assess the merit of a target set of $n$ scientific papers, by calculating the probability that a randomly selected set of $n$ articles from a given pool of articles in the same field has a lower citation impact indicator than the target set, where the citation impact indicator can be the mean citations or the h-index. Although their approach is quite different from ours (and they do not include papers published in multidisciplinary journals), the underlying motivation is similar, as it is based on the realization that both the mean citations and the h-index are actually not independent of size. 

Thus, counter-intuitively, perhaps
(Waltman, 2016),
the process of averaging does not guarantee size independence. But then, what is the point in using averages? Should we avoid them altogether? Of course not. When there is low disparity (variance) within the population, or when sample sizes are large enough to absorb the fluctuations, then averages are fine. But in citation wealth, as in actual wealth, the vast disparity within the population makes plain (i.e, unnormalized) averages misleading, especially when very small samples are involved. After all, we are used to reports of average wealth (GDP per capita) for nations (large populations), but median house prices for neighborhoods (small populations).

\section{But actual journals are not random paper selections...}
\label{S:6}

The Central Limit Theorem is based on the sampling distribution, in a process of randomly drawing $n$ papers from the entire scientific literature and `forming' a journal, many times over, and then studying the statistics of the citation averages produced. Even though it may appear counterintuitive at first, we have established in this paper that the Central Limit Theorem successfully explains and quantifies many properties of the scale-dependent behavior of Impact Factors. But surely there are other aspects of Impact Factors that cannot be attributed to randomness and the Theorem? Indeed there are---and to find them, we need to look for {\it sustained above-average performance}, where  ``above average"  is meant ``with respect to $\mu$.'' If a journal repeatedly (i.e., year after year) scores a ``high IF"---close to $f_{max}^{th}$, or even higher---then this behavior cannot be explained by chance and the Central Limit Theorem. An alternative hypothesis must be at work, where the journal's high IF results from the concerted efforts by authors, editors, editorial boards, and referees, to uphold high standards. 
(At least, this is the case for journals in similar research fields. For as we stated in {\it implication \#2} of \S \ref{SS:2.2}, we do not differentiate here for different citation practices across research fields.) 
Indeed, there are many journal datapoints in Fig. \ref{fig:JIF-NY_alljnls_clean} whose IFs deviate substantially {\it and} repeatedly from the Central Limit Theorem predictions. For example,  2576 out of 136874 IF \& journal-size data pairs in 1997--2016 with annual size $N_Y>25$, had their ${\rm IF} >f_{max}^{th}$. For 509 journal datapoints, ${\rm IF} >f_{max}^{th}+10$. We already encountered such journals in \S \ref{SSS:4.2.3}, when we noticed  local peaks (groups of high-IF datapoints) at journal sizes $N_Y \approx 350$, where titles such as the {\it New England Journal of Medicine, Lancet, Chemical Reviews, Journal of the American Medical Association, Cell, Nature Reviews Molecular Cell Biology, Nature Materials, Nature Nanotechnology,} etc., routinely scored well above $f_{max}^{th}$ at the corresponding size. Another example was at $N_Y \approx 900$, where {\it Nature} and {\it Science} also score well above $f_{max}^{th}$ year after year. In addition, there is an even higher number of journals that {\it repeatedly} `float' below yet near the $f_{max}^{th}$ curve, and this is no statistical accident either: Even though the Central Limit Theorem predicts that IFs of randomly selected journals will range from $f_{min}^{th}$ to $f_{max}^{th}$ for a given size $n$, if the journals were truly random selections of papers they would each `scan' all the IF range of values available to them, given enough time. And yet many journals consistently stay near $f_{max}^{th}$. Again, this can only be the result of a collective effort by several `actors' for  high-quality content.  

At the other end of the spectrum, if a journal has an ${\rm IF}  < \mu$, then it demonstrates below-average performance (``average" with respect to $\mu$). And if its IF is consistently below $\mu$ and close to (or even lower than) the theoretical minimum, $f_{min}^{th}$, then we can say that the journal is  consistently not attracting or  retaining well-cited papers.

\section{The $\Phi$ index: A scale-adjusted Impact Factor}
\label{S:7}

The above discussion  leads us naturally to consider an alternative measure of scale-adjusted citation impact. Let us define the $\Phi$ index,
\begin{equation}
\Phi =  \frac{{\rm IF}-\mu}{f_{max}^{th}(n)-\mu}. \label{eq:Phi} 
\end{equation}
Defined this way, $\Phi$ is a measure of a journal's citation impact offset by the average $\mu$ and normalized to the maximum IF uncertainty $\Delta f_{max}=f_{max}^{th}(n)-\mu$ (see Eq. (\ref{eq:IFUP.3})). Clearly, $\Phi=0$ for average performance, $\Phi>0$ for above-average performance, and $\Phi<0$ for below-average performance.  The normalization is chosen so that journals with  ${\rm IF}=f_{max}^{th}(n)$ have $\Phi=1$, while journals with  ${\rm IF}=f_{min}^{th}(n)$ have $\Phi=-1$ (recall that $f_{max}^{th}(n)$ and $f_{min}^{th}(n)$ are symmetric about $\mu$). Journals that are ``equidistant'' from the average $\mu$, where the distance is {\it normalized} to $\Delta f_{max}\equiv f_{max}^{th}(n)-\mu$, are ranked equal. 

We can substitute  $f_{max}^{th}$ from Eq. (\ref{eq:fmax-fmin}) into Eq. (\ref{eq:Phi}), to obtain  
\begin{equation}
\Phi =  \frac{({\rm IF}-\mu)\sqrt[]{n}}{k \sigma}, \label{eq:Phi2}
\end{equation}
which can be readily calculated for journals once $\mu$ and $\sigma$ are known or at least estimated. If one is merely interested in the relative ranks of journals, one can drop the denominator $k \sigma$ and compare the values $({\rm IF}-\mu)\sqrt[]{n}$, where $\mu \approx 3$. (The underlying assumption for such a comparison is that journals belong to the same population, in the sense described in {\it implication \#2} of \S \ref{SS:2.2}. One can make this assumption as a zeroth-order approximation, but for more detailed comparisons, the distinct $\sigma$ and $\mu$ for each population should be used as necessary. Indeed, some sort of  field normalization, or accounting for citation differences across different populations (research fields) of  papers, is important for Impact Factors---and therefore also for the $\Phi$ index.)

In Table 2 we list the top-50 journals, ranked by their 2016 $\Phi$ index. In the table, we also list the IF value, the journal biennial size $N_{2Y}$, and the IF rank. We observe that 31 journals in the top-50 $\Phi$ ranks are also in the top-50 IF ranks. Of the remaining 19, the most interesting new entries (compared to the top-50 IF ranks) are: J. Am. Chem. Soc. (\#10 from \#111), Nat. Commun. (\#12 from \#148), Angew. Chem. Int. Edit. (\#16 from \#150), ACS Nano (\#20 from \#109), PNAS (\#22 from \#210), Nano Lett. (\#28 from \#133), J. Mater. Chem. A (\#31 from \#249), Blood (\#33 from \#125), Phys. Rev. Lett. (\#38 from \#273),  Nucleic Acids Res. (\#39 from \#196), Adv. Funct. Mater.  (\#44 from \#147), and ACS Appl. Mater. Interf. (\#46 from \#327). All these journals are scoring well above average {\it given their size}. As for the three megajournals, their $\Phi$ ranks, compared to their IF ranks, are as follows: Sci. Rep., \#174 from \#973; RSC Adv., \#3445 from \#1888; and {\it PLoS ONE}, \#8811 from \#2264.  So, Sci. Rep. is pulled to higher rank by the $\Phi$ index compared to the IF, while RSC Adv. and {\it PLoS ONE} are pushed to lower ranks. This happens because the IFs of RSC Adv. and {\it PLoS ONE} are below average (at 3.108 and 2.806 respectively, compared to $\mu=3.2$) while the IF of Sci. Rep. (4.259) is above average. 

Thus the definition of the $\Phi$ index is an attempt to correct for two unfair advantages that result from the Central Limit Theorem: (a) the advantage of small journals to score a high-IF---in which case, $({\rm IF}-\mu)$ is high but  $n$ is low---and (b) the advantage of megajournals to score an IF close to the population average $\mu$---in which case, $n$ is high but $({\rm IF}-\mu)$ is low. See Eqs. (\ref{eq:Phi}, \ref{eq:Phi2}). A more detailed analysis of the $\Phi$ index will be presented in a forthcoming publication.

\begin{table}
\centering
\begin{tabular}{lrrrr}
\hline
Journal (ranked by $\Phi$ index) & IF & $\Phi$ & $N_{2Y}$ & IF rank\\
\hline
1.	NEW ENGL J MED	&	72.406	&	{\bf	18.24	} &	695	&	2	\\
2.	NATURE	&	40.137	&	{\bf	15.51	} &	1763	&	10	\\
3.	SCIENCE	&	37.205	&	{\bf	13.84	} &	1656	&	16	\\
4.	CA-CANCER J CLIN	&	187.040	&	{\bf	13.00	} &	50	&	1	\\
5.	LANCET	&	47.831	&	{\bf	10.75	} &	580	&	5	\\
6.	CHEM REV	&	47.928	&	{\bf	10.41	} &	542	&	4	\\
7.	CHEM SOC REV	&	38.618	&	{\bf	9.63	} &	740	&	14	\\
8.	JAMA-J AM MED ASSOC	&	44.405	&	{\bf	8.49	} &	425	&	7	\\
9.	CELL	&	30.410	&	{\bf	8.04	} &	873	&	22	\\
10.	J AM CHEM SOC	&	13.858	&	{\bf	7.56	} &	5030	&	111	\\
11.	ADV MATER	&	19.791	&	{\bf	7.33	} &	1954	&	58	\\
12.	NAT COMMUN	&	12.124	&	{\bf	6.91	} &	6001	&	148	\\
13.	ENERG ENVIRON SCI	&	29.518	&	{\bf	6.86	} &	680	&	24	\\
14.	NAT MATER	&	39.737	&	{\bf	6.57	} &	323	&	12	\\
15.	J CLIN ONCOL	&	24.008	&	{\bf	6.49	} &	973	&	39	\\
16.	ANGEW CHEM INT EDIT	&	11.994	&	{\bf	6.24	} &	5035	&	150	\\
17.	NAT NANOTECHNOL	&	38.986	&	{\bf	6.14	} &	294	&	13	\\
18.	NAT BIOTECHNOL	&	41.667	&	{\bf	5.77	} &	225	&	8	\\
19.	LANCET ONCOL	&	33.900	&	{\bf	5.66	} &	340	&	19	\\
20.	ACS NANO	&	13.942	&	{\bf	5.47	} &	2596	&	109	\\
21.	NAT PHOTONICS	&	37.852	&	{\bf	5.40	} &	243	&	15	\\
22.	P NATL ACAD SCI USA	&	9.661	&	{\bf	5.36	} &	6870	&	210	\\
23.	NAT GENET	&	27.959	&	{\bf	4.87	} &	387	&	29	\\
24.	NAT REV DRUG DISCOV	&	57.000	&	{\bf	4.78	} &	79	&	3	\\
25.	J AM COLL CARDIOL	&	19.896	&	{\bf	4.76	} &	814	&	56	\\
26.	NAT MED	&	29.886	&	{\bf	4.75	} &	317	&	23	\\
27.	NAT REV MOL CELL BIO	&	46.602	&	{\bf	4.71	} &	118	&	6	\\
28.	NANO LETT	&	12.712	&	{\bf	4.62	} &	2363	&	133	\\
29.	CIRCULATION	&	19.309	&	{\bf	4.61	} &	818	&	61	\\
30.	ACCOUNTS CHEM RES	&	20.268	&	{\bf	4.39	} &	663	&	49	\\
31.	J MATER CHEM A	&	8.867	&	{\bf	4.13	} &	5309	&	249	\\
32.	NAT REV IMMUNOL	&	39.932	&	{\bf	3.99	} &	118	&	11	\\
33.	BLOOD	&	13.164	&	{\bf	3.96	} &	1581	&	125	\\
34.	EUR HEART J	&	19.651	&	{\bf	3.95	} &	576	&	60	\\
35.	NAT METHODS	&	25.062	&	{\bf	3.93	} &	323	&	38	\\
36.	BMJ-BRIT MED J	&	20.785	&	{\bf	3.91	} &	494	&	47	\\
37.	NAT REV GENET	&	40.282	&	{\bf	3.89	} &	110	&	9	\\
38.	PHYS REV LETT	&	8.462	&	{\bf	3.83	} &	5287	&	273	\\
39.	NUCLEIC ACIDS RES	&	10.162	&	{\bf	3.69	} &	2813	&	196	\\
40.	NAT REV CANCER	&	37.147	&	{\bf	3.66	} &	116	&	17	\\
41.	CANCER CELL	&	27.407	&	{\bf	3.64	} &	226	&	30	\\
42.	NAT CHEM	&	25.870	&	{\bf	3.61	} &	253	&	36	\\
43.	ADV ENERGY MATER	&	16.721	&	{\bf	3.55	} &	691	&	79	\\
44.	ADV FUNCT MATER	&	12.124	&	{\bf	3.55	} &	1583	&	147	\\
45.	IMMUNITY	&	22.845	&	{\bf	3.53	} &	323	&	41	\\
46.	ACS APPL MATER INTER	&	7.504	&	{\bf	3.36	} &	6112	&	327	\\
47.	GASTROENTEROLOGY	&	18.392	&	{\bf	3.35	} &	485	&	65	\\
48.	NAT PHYS	&	22.806	&	{\bf	3.30	} &	284	&	43	\\
49.	LANCET NEUROL	&	26.284	&	{\bf	3.12	} &	183	&	35	\\
50.	NAT NEUROSCI	&	17.839	&	{\bf	3.08	} &	442	&	69	\\
\hline
\end{tabular}
\caption{Journal rankings by  $\Phi$ index. Rankings by Impact Factor shown in  last column.}
\end{table}



\section{Conclusions and outlook}

In this paper, we have used the Central Limit Theorem of statistics to understand the behavior of citation averages (Impact Factors). We find that Impact Factors are strongly dependent on journal size. We explain the observed stratification of journals in Impact Factor rankings, whereby small journals occupy the top, middle, {\it and} bottom ranks; mid-sized journals occupy the middle ranks; and very large journals (``megajournals") converge to a single Impact Factor value. Further, we applied the Central Limit Theorem to develop an uncertainty relation for Impact Factors in the form of Eq. (\ref{eq:IFUP.2}), a mathematical expression for the range of IF values statistically available to journals of a certain size. We confirm the functional form of the upper bound of the IF range by analyzing the complete set of 166,498 IF and journal-size data pairs in the 1997--2016 Journal Citation Reports (JCR) of Clarivate Analytics, the top-cited portion of 276,000 physics papers published in 2014--2015, as well as the citation distributions of an arbitrarily sampled list of physics journals. 

The Central Limit Theorem effects are strong enough to interfere with IF values and affect the corresponding rankings. The Impact Factor uncertainty relation quantifies these effects as a function of any journal size, but the two main unfair advantages we have identified are that (a) small journals can attain very high IFs, while (b) very large journals will avoid low IFs. The former advantage is unfair towards all mid-sized and large journals, the latter is unfair towards low-IF small and mid-sized journals. It is thus suggested to compare like-sized journals in IF rankings. If one {\it must} compare different-sized journals, it would be better to use the rescaled index $\Phi=({\rm IF}-\mu)/(f_{max}^{th}(n)-\mu)$. 

One could analyze the uncertainty relation Eq. (\ref{eq:IFUP.2}) by research fields, which would result in field-specific $\mu$ and $\sigma$, and likewise for the upper and lower bounds for Impact Factors. 
Field normalization (or accounting for citation differences across fields one way or another) is standard procedure for citation averages. Our objective is to introduce size-normalization---expressed via the $\Phi$ index---{\it in addition} to field normalization of citation averages.

It would also be interesting to study the effect of citation inflation, which we have ignored here, on $\mu$ and $\sigma$ for the citations population of research papers, even though it is reasonable to expect that this would be rather small 
(Althouse et al., 2009).

We have applied the Central Limit Theorem to understand Impact Factors. But there is nothing exclusive about journals in our analysis, and the same methodology can be applied to other types of citation averages, to describe university departments, entire universities, or even countries, as in various global rankings. The key features to keep in mind when assessing such rankings of average quantities, are that (a) the Central Limit Theorem allows small entities to fluctuate much more vividly and thus reach much higher values than larger entities; and (b) very large entities will converge to a single value, characteristic of the population itself.  

Before closing, let us briefly comment on the size-dependence of indicators other than arithmetic averages, such as percentile-based indicators (Leydesdorff \& Bornmann, 2011) or geometric averages (Thelwall \& Fairclough, 2015). Because the Central Limit Theorem applies specifically to arithmetic averages, the $1/\sqrt[]{n}$  dependence that we observe here does not apply to percentile-based indicators or geometric averages. An analysis of size effects for such indicators would be interesting but is clearly beyond the scope of this study. Let us simply register an observation, drawn from the physics papers published in 2014--2015 and cited in 2016. For this set of papers, the range of values of the {\it geometric} average---determined by its maximum value---decreases with sample size (i.e., decreasing citation rank) $n$ as $\sim n^{-0.225}$. While this decrease is slower that for the corresponding arithmetic average of Eq. (\ref{eq:f_max=k^-0.44})---and, curiously, with exactly {\it half} the exponent---it still suggests that some form of size-normalization if also necessary for geometric averages of citation distributions.

\vspace{1cm}

{\it Acknowledgments}: I am grateful to Jerry I. Dadap (Columbia University), Albyn Jones (Reed College), Andrew Gelman (Columbia University), and Amine Triki (Clarivate Analytics) for stimulating discussions, and to Richard Osgood Jr. (Columbia University) for encouragement and hospitality. 

\vspace{0.5cm}
{\it Disclaimer: The author is an Associate Editor at the American Physical Society. The opinions expressed here are his own.}

\vspace{0.5cm}
This research did not receive any specific grant from funding agencies in the public, commercial, or not-for-profit sectors.

\subsection*{References}
\vspace{0.5cm}

\noindent 
Adler, R., Ewing, J., Taylor, P., (2008). Citation Statistics. {\it Joint Committee on Quantitative 
\par 
Assessment of Research}, available at 
\par 
http://www.mathunion.org/fileadmin/IMU/Report/CitationStatistics.pdf. 

\noindent 
Althouse, B. M., West, J. D., Bergstrom, C. T., Bergstrom, T., (2009). Differences in Impact 
\par 
Factor across fields and over
time, {\it Journal of the American Society for Information Science 
\& Technology, 60}, 27--34.

\noindent 
Amin, M., Mabe, M., (2004). Impact factors: Use and abuse, {\it International Journal of Environ-
\par 
mental Science and Technology, 1}, 1--6.

\noindent 
Antonoyiannakis, M., (2008). Citation analysis: Beyond the Journal Impact Factor. {\it APS March 
\par 
Meeting}, Available at
http://meetings.aps.org/link/BAPS.2008.MAR.U39.8.

\noindent 
Antonoyiannakis, M., (2015a). Impact Facts \& Impact Factors: A view from the Physical 
\par 
Review.
{\it PQE - The 45th Winter Colloquium on the Physics of Quantum Electronics,} 
\par 
Available at
http://www.pqeconference.com/pqe2015/Abstracts/345p.pdf. 

\noindent 
Antonoyiannakis, M., (2015b). Median Citation Index vs Journal Impact Factor. {\it APS March 
\par 
Meeting}, Available at
http://meetings.aps.org/link/BAPS.2015.MAR.Y11.14.

\noindent 
Antonoyiannakis, M., Mitra, S., (2009). Editorial: Is PRL too large to have an `impact'? {\it Physical 
\par 
Review Letters, 102},
060001.

\noindent 
Bornmann, L., Leydesdorff, L., (2017). Skewness of citation impact data and covariates of 
\par 
citation distributions: A large-scale empirical analysis based on Web of Science data. 
\par
{\it Journal of Informetrics, 11}, 164--175.

\noindent 
Bornmann, L.,  Leydesdorff, L. (2018). Count highly-cited papers instead of papers with h 
\par
citations: use normalized citation counts and compare ``like with like''!, {\it Scientometrics} 115, 
\par
1119--1123.

\noindent 
Crespo, J., Ortu{\~n}o-Ort{\'i}n, I., Ruiz-Castillo, J., (2012). The citation merit of scientific 
\par publications, {\it PLoS ONE, 7}, e49156. 

\noindent 
De Veaux, R. D., Velleman, P. D., \& Bock, D. E. (2014). Stats: Data and Models (3rd Edition), 
\par {\it Pearson Education Limited.}

\noindent
Fang, L., Wenbo, G., Chao Z., (2018). High impact factor journals have more publications 
\par 
than expected. {\it Current Science, 114}, (5), 955--956.

\noindent 
Fersht, A., (2009). The most influential journals: Impact Factor and Eigenfactor. {\it Proceedings of 
\par the National Academy of Sciences of the United States of America, 106}, 6883--6884.

\noindent 
Gaind, N. (2018). Few UK universities have adopted rules against impact-factor abuse, {\it Nature 
\par News}, Available at
https://www.nature.com/articles/d41586-018-01874-w.

\noindent 
Gelman, A., Nolan, D., (2002). Teaching Statistics: A Bag of Tricks, {\it Oxford University Press.}

\noindent 
Gl{\"a}nzel, W., Moed, H. F., (2013) Opinion paper: Thoughts and facts on bibliometric indicators. 
\par {\it Scientometrics, 96}, 381--394.

\noindent 
Huang, D.-w., (2016). Positive correlation between quality and quantity in academic journals, 
\par {\it Journal of Informetrics 10}, 329--335.

\noindent 
Katz, J.S., (2000). Scale-independent indicators and research evaluation, {\it Science and Public 
\par Policy, 27}, 23--36.

\noindent 
Leydesdorff, L., \& Bornmann, L., (2011), Integrated Impact Indicators (I3) compared with 
\par
Impact Factors (IFs): An alternative design with policy implications, {\it Journal of the American 
\par
Society for Information Science and Technology, 62} (11), 2133--2146.

\noindent 
Merton, R., (1968). The Matthew effect in science, {\it Science, 159,} 56--63.

\noindent 
Moed, H. F. (2005). Citation Analysis in Research Evaluation, {\it Springer}.

\noindent 
Newman, M. E. J., (2005). Power laws, Pareto distributions and Zipf's law, {\it Contemporary 
\par Physics, 46}, (5), 323--351.

\noindent 
Radicchi, F., Fortunato, S., Castellano, C., (2008). Universality of citation distributions: Toward 
\par an objective measure of
scientific impact. {\it Proceedings of the National Academy of Sciences 
\par of the United States of America, 105}, 17268--17272.

\noindent 
Redner, S., (1998). How popular is your paper? An empirical study of the citation distribution. 
\par {\it European Physical Journal B, 4}, 131--134.

\noindent 
Rossner, M., Van Epps, H., Hill, E., (2007). Show me the data. {\it Journal of Cell Biology 179}, 
\par 1091--1092.

\noindent 
Rousseau, R. (2016). Positive correlation between journal production and journal impact factor. 
\par
{\it Journal of Informetrics, 10}, 567--568. 

\noindent 
Rousseau, R., Van Hooydonk, G., (1996). Journal production and journal impact factors, {\it Journal 

of the American Society for Information Science, 47}, 775--780.

\noindent 
San Francisco Declaration on Research Assessment, (2012). Available at 
\par http://www.ascb.org/dora/.

\noindent 
Seglen, P. O., (1992). The skewness of science. {\it Journal of the American Society for Information 
\par Science, 43}, 628--638. 

\noindent 
Seglen, P. O., (1997). Why the Impact Factor of journals should not be used for evaluating 
\par research. {\it British Medical Journal, 314}, 498--502. 

\noindent 
Stephan, P., Veugelers, R., Wang, J., (2017). Reviewers are blinkered by bibliometrics. {\it Nature, 
\par 544}, 411--412.

\noindent 
Thelwall, M., Fairclough, R., (2015). Geometric journal impact factors correcting for individual 
\par
highly cited articles. {\it Journal of Informetrics, 9}, 263--272.

\noindent 
van Raan, A. F., (2008). Bibliometric statistical properties of the 100 largest European research 
\par universities: Prevalent
scaling rules in the science system, {\it Journal of the American Society 
\par for Information Science \& Technology, 59}, 461--475.

\noindent 
Wainer, H., Zwerling, H., (2006). Legal and empirical evidence that smaller schools do not 
\par improve student achievement.
{\it The Phi Delta Kappan, 87}, 300--303.

\noindent 
Wainer, H., (2007). The most dangerous equation, {\it American Scientist,} May-June 249--256.

\noindent 
Wall, H. J., (2009). Don't get skewed over by journal rankings. {\it The B.E. Journal of Economic 
\par Analysis \& Policy, 9}, article 34.

\noindent 
Waltman, L., Calero-Medina, C., Kosten, J., Noyons, E.C.M.,  Tijssen, R.J.W., van Eck, N.J., 
\par
 van Leeuwen, T.N.,  van Raan, A.F.J.,  Visser, M.S., and Wouters, P. (2012). The Leiden 
\par 
Ranking 
2011/2012: Data Collection, Indicators, and Interpretation, {\it Journal of the 
\par
American Society of Information Science and Technology}, {\it 63}(12), 2419--2432.

\noindent 
Waltman, L., (2016). A review of the literature on citation impact indicators, {\it Journal of 
\par Informetrics, 10}, 365--391.











\bibliographystyle{model1-num-names}

\end{document}